\documentclass[journal]{IEEEtran}
\usepackage{cite}
\usepackage{graphicx}
\usepackage{subfigure}
\usepackage{url}
\usepackage{booktabs}
\usepackage{epsfig}
\usepackage{amsmath}
\usepackage{amsfonts}
\usepackage{amssymb}
\usepackage{mathrsfs}
\usepackage[final,scrtime]{prelim2e}  
\usepackage{verbatim}
\usepackage{algorithm}
\usepackage{algorithmic}
\usepackage{ntheorem}
\usepackage{url}
\usepackage{color}
\usepackage{multirow}     
\usepackage{diagbox}    
\usepackage{array}      
\usepackage{caption}    
\usepackage{colortbl}
\usepackage{longtable}   
\usepackage{xcolor}
\usepackage{xpatch}
\usepackage{lipsum,mwe,cuted}
\usepackage{stfloats}
\usepackage{enumerate}  
\usepackage{amsmath,mathtools}
\usepackage{makecell}

\graphicspath{{./}}
\theorembodyfont{\upshape}
\theoremheaderfont{\rmfamily\itshape}
\theoremseparator{:}

\newtheorem*{pproof}{\hspace{1em}Proof}
\theoremstyle{remark}
\newtheorem{thm}{\hspace{1em}Theorem}
\newtheorem{cor}{\hspace{1em}Corollary}
\newtheorem{lem}{\hspace{1em}Lemma}

\newcommand{\tabincell}[2]{\begin{tabular}{@{}#1@{}}#2\end{tabula}}

\theoremstyle{remark}

\makeatletter
\def\changeBibColor#1{%
\in@{#1}{}
\ifin@\color{red}\else\normalcolor\fi
}
\xpatchcmd\@bibitem
{\item}
{\changeBibColor{#1}\item}
{}{\fail}
\xpatchcmd\@lbibitem
{\item}
{\changeBibColor{#2}\item}
{}{\fail}
\makeatother
\definecolor{r}{rgb}{1,0,0}
\definecolor{b}{rgb}{0,0,1}
\definecolor{k}{rgb}{0,1,1}

\def\mb{\mathbf}

\def\rm{\textrm}

\renewcommand{\vec}{\mb}    

\begin{document}
\title{A Novel Cost-Effective MIMO Architecture with Ray Antenna Array for Enhanced Wireless Communication Performance}
\author{
Zhenjun Dong, Zhiwen Zhou, Yong Zeng,~\IEEEmembership{Fellow,~IEEE}
        \thanks{Part of this work will be presented at IEEE VTC2025 Spring, Oslo, Norway, 2025 \cite{zhenjunRAA}.}
       \thanks{Z. Dong, Z. Zhou, and Y. Zeng are with the National Mobile Communications Research Laboratory, Southeast University, Nanjing 210096, China. Y. Zeng is also with the Purple Mountain Laboratories, Nanjing 211111, China (e-mail: \{zhenjun\_dong, zhiwen\_zhou, yong\_zeng\}@seu.edu.cn). (Corresponding author: Yong Zeng.)}

}
\maketitle
\begin{abstract}
This paper proposes a novel multi-antenna architecture, termed ray antenna array (RAA), which practically enables  flexible beamforming and also enhances wireless communication performance for high-frequency systems in a cost-effective manner.
RAA consists of a large number of inexpensive antenna elements and a few radio frequency (RF) chains.
These antenna elements are arranged in a novel ray-like structure, where each ray corresponds to one \emph{simple uniform linear array} (sULA) with a carefully designed orientation.
The antenna elements within each sULA are directly connected, so that each sULA is able to form a beam towards a direction matching the ray orientation without relying on any analog or digital beamforming.
By further designing a ray selection network (RSN), appropriate sULAs are selected to connect to the RF chains for subsequent baseband processing.
Compared to conventional multi-antenna architectures such as the uniform linear array (ULA) with hybrid analog/digital beamforming (HBF), the proposed RAA enjoys three appealing  advantages: (i) finer and uniform angular resolution for all signal directions; (ii) enhanced beamforming gain by using antenna elements with higher directivity, as each sULA is only responsible for a small portion of the total angle coverage range;
and (iii)
dramatically reduced hardware cost since no phase shifters are required, which are expensive and difficult to design in high-frequency systems such as millimeter wave (mmWave) and Terahertz (THz) systems.
To validate such advantages, we first present the input-output mathematical model for RAA-based wireless communications. 
Efficient algorithms for joint RAA beamforming and ray selection are then proposed for single-user and multi-user RAA-based wireless communications.
Simulation results demonstrate that RAA achieves superior performance compared to the conventional ULA with HBF, while significantly reducing hardware cost.
\end{abstract}

\section{Introduction}
The continuous advancement of multi-antenna or multiple-input multiple-output (MIMO) technology has been pivotal to the evolution of wireless communication systems.
To fully exploit the spatial degrees of freedom (DoFs), the scale of antenna array has expanded considerably, evolving from traditional MIMO with typically 8 antennas in the fourth-generation (4G) mobile communication networks, to massive MIMO with typically 64 antennas in the fifth-generation (5G) networks \cite{heath2018foundations}.
The forthcoming sixth-generation (6G) mobile communication networks are envisioned to not only further significantly  improve communication performance, but also provide high-quality wireless sensing and localization services \cite{xiao2022overview}.
This brings new opportunities and design challenges for the further advancement of MIMO technology.
To this end, various new MIMO technologies
 such as near-field extremely large-scale MIMO (XL-MIMO) \cite{10496996}, cell-free massive MIMO \cite{ngo2017cell}, and sparse massive MIMO \cite{li2024sparse} are being actively explored.
There is 
a general consensus that 6G will require larger antenna arrays, higher frequency bands, and wider spectrum resources.
However, indiscriminately scaling system dimensions across spatial and time/frequency domains may exacerbate the energy consumption and operational costs, potentially conflicting with other critical 6G design objectives such as improving energy efficiency and reducing cost.
Consequently, the development of advanced multi-antenna technology that can sustain or even enhance system performance while maintaining economic viability is of paramount importance.

Since radio frequency (RF) chains and front-end modules account for a substantial portion of hardware costs and energy consumption, an effective method to reduce hardware and energy costs for MIMO systems is to use fewer RF chains or employ more cost-effective RF components \cite{zeng2024tutorial}.
To this end, analog beamforming or phased antenna array \cite{wang2009beam,hur2013millimeter} has been extensively studied, where flexible beams are formed by dynamically adjusting the phases of RF signals using analog components.  However, analog beamforming is inherently limited to single-stream MIMO transmission.
An alternative approach, hybrid analog/digital beamforming (HBF) \cite{el2014spatially,heath2016overview,mendez2016hybrid,gao2016energy},
provides a more flexible trade-off between cost and performance by using fewer RF chains than antenna elements.
Based on the connectivity between RF chains and antennas, HBF architectures are typically categorized into fully-connected \cite{el2014spatially,heath2016overview} and partially-connected structures \cite{mendez2016hybrid,gao2016energy}. The partially-connected architecture requires fewer phase shifters than the fully-connected counterpart, but it compromises the effective array aperture.
In fact, for both partially and fully-connected HBF architectures, the number of phase shifters required increases significantly with the number of antenna elements.
This makes it more challenging for the implementation of XL-array in 6G, 
 as the cost and complexity of analog phase shifters can quickly become unsustainable.
Particularly for high-frequency systems, such as millimeter-wave (mmWave) and Terahertz (THz) systems, achieving accurate phase shifter design presents considerable technical difficulties due to factors such as high insertion loss, limited phase resolution, and sensitivity to process variations \cite{ahmed2018survey,han2021hybrid,pang201928}.
The implementations of phase shifters, such as loaded-line and varactor-based designs, further suffer from severe parasitic effects and power-handling constraints. Thus, the miniaturization of phase shifter components at high carrier frequency exacerbates fabrication complexities, posing significant challenges to the cost-effective realization of XL-MIMO.


To address the above challenges and further reduce hardware costs, researchers are actively exploring various new  methods, such as switch-based HBF architecture \cite{mendez2016hybrid,7370753,payami2018phase,qi2022hybrid}, lens antenna array \cite{zeng2016millimeter}, and the recently proposed tri-hybrid architecture based on configurable antenna array \cite{castellanos2025embracing}.
In switch-based HBF architectures, some works proposed completely replacing costly phase shifters with low-cost RF switches\cite{mendez2016hybrid,7370753}, while others combined both switches and phase shifters\cite{payami2018phase,qi2022hybrid}. However, switch-based implementations often suffer from degraded system performance due to their coarse phase granularity and binary on/off control, thus limiting fine-grained beam steering.
In particular, switch-based architectures may lead to significant array aperture loss when only a subset of antennas is activated at each time, thereby reducing effective array gain and spatial resolution.
In \cite{zeng2016millimeter}, an electromagnetic lens-enabled array structure was developed, which can concentrate the energy of incident signals onto a small area of the antenna array, thereby reducing hardware costs and power consumption.
Recently, a tri-hybrid architecture that combines HBF with reconfigurable antenna arrays has been proposed \cite{castellanos2025embracing}, and some representative reconfigurable antenna architectures include fluid antenna (FA) \cite{wong2020fluid,10753482}, movable antenna (MA) \cite{zhu2023movable,10906511,dong2024movable}, and group MA \cite{lu2024group}.
While the tri-hybrid architectures offer enhanced beamforming flexibility, they typically require a large number of reconfigurable antenna elements. 
Moreover, by shifting part of signal processing to the antenna domain, the tri-hybrid architecture necessitates sophisticated antenna-side hardware and localized processing capabilities. These factors significantly increase the overall implementation cost and energy overhead, rendering them not easy for scalable and energy-efficient 6G deployments.



To enable flexible beamforming and further improve wireless performance particularly for high-frequency systems such as mmWave and THz systems, we propose a novel multi-antenna architecture, termed {\it ray antenna array} (RAA).
The RAA architecture is inspired by the observation that even without any analog or digital beamforming, a so-called {\it simple uniform linear array} (sULA) with all antenna elements directly connected can still form a beam in the direction matching its physical orientation.
By deploying multiple such sULAs with deliberately designed orientations, flexible beamforming towards the desired directions can be achieved.
As a result, RAA provides a beamforming capability without requiring any phase shifters, which can substantially reduce hardware costs.
Although RAA requires a larger number of antenna elements, they are typically low-cost, and their integration is facilitated by the short wavelength in high-frequency bands. This makes RAA particularly suitable for mmWave and THz systems, where dense packaging of antenna elements is feasible.
The three key contributions of this paper are summarized as follows.
\begin{itemize}

\item
First, we present the detailed design of the proposed RAA architecture.
To characterize the behavior of the RAA, we develop the array response model of a sULA with any given orientation, which captures the geometric structure of RAA and the radiation pattern of individual antenna elements.
Based on this model, we derive the beam pattern of each sULA as a function of the array orientation and signal direction.
By analyzing the beam pattern of a given sULA, we establish a design criterion for determining the orientation of each sULA. Considering both the orientation criterion and the total angle of arrival/departure (AoA/AoD) coverage range, we determine the required number of sULAs. 
Furthermore, by leveraging the structural characteristics of the RAA architecture, we rigorously prove the inherent directivity of individual RAA antenna elements. 
Compared to the conventional ULA, the proposed RAA enjoys three significant advantages in terms of angular resolution, overall beamforming gain, and hardware cost:
  (i) \textbf{Finer and uniform angular resolution}: 
  Compared to the classic ULA with the same array gain factor, RAA achieves finer and uniform angular resolution for all signal directions; 
  (ii) \textbf{Enhanced beamforming gain}: RAA can provide enhanced overall beamforming gain since antenna elements with higher directivity can be used, considering that each sULA is only responsible for a portion of the total angle coverage range;
  (iii) \textbf{Dramatically reduced hardware cost}: By eliminating the need for phase shifters, which are costly and technically challenging to design especially in mmWave and THz systems, RAA can significantly reduce hardware costs.

\item
Next, we propose the wireless communication system based on the designed RAA architecture.
To enable flexible beam control, a ray selection network (RSN) is introduced, which dynamically selects the most appropriate sULAs to connect to the RF chains for subsequent baseband processing.
Furthermore, we propose a joint RAA beamforming and ray selection algorithm for both uplink and downlink scenarios in the RAA-based communications.
In the special case of single-user communication, the design for uplink and downlink is unified by maximizing the signal-to-noise ratio (SNR) through maximum ratio combining/transmission (MRC/MRT) in baseband, and selecting the sULAs  with the strongest effective channel responses.
For the uplink multi-user case, we propose an efficient greedy algorithm to maximize the communication sum rate. Compared to exhaustive search over all possible sULA combinations, the proposed greedy algorithm achieves comparable performance with significantly reduced complexity.
For downlink communication, we propose an alternating optimization algorithm that jointly designs RAA beamforming and ray selection to maximize the minimum signal-to-interference-plus-noise ratio (SINR) among all users. 

\item
Last, extensive simulation results are provided to validate the effectiveness of the proposed RAA-based system.
The simulations show that the RAA architecture significantly outperforms conventional ULA architecture with HBF. Specifically, simulation results verify that the proposed RAA achieves finer and uniform angular resolution compared to the conventional ULA with the same array gain factor. 
Moreover, the RAA system yields notable improvements in communication performance for both uplink and downlink scenarios, benefiting from enhanced beamforming gain and spatial resolution.
In addition, the simulation results confirm the effectiveness of the proposed greedy algorithm for uplink sum-rate maximization and the alternating optimization algorithm for downlink max-min SINR, both of which achieve near-optimal performance with substantially reduced complexity.
\end{itemize}

The remainder of the paper is organized as follows.
The uplink multi-user system model based on the proposed RAA architecture is presented in Section II.
The mathematical modeling and parameter design of the RAA architecture are detailed in Section III.
Section IV focuses on the design of joint RAA beamforming and ray selection for uplink multi-user communications.
Section V extends the study to downlink RAA-based multi-user communications.
Simulation results are discussed in Section VI. Finally, we conclude this paper in Section VII.

{\it Notation}:
Italic, bold-faced lower- and upper-case letters denote scalars, vectors, and matrices, respectively.
The transpose, Hermitian transpose, and complex conjugate operation are given by $(\cdot)^T$, $(\cdot)^H$, and $(\cdot)^{*}$, respectively.
$\mathbb{C}^{M\times N}$ and $\mathbb{R}^{M\times N}$ signify the spaces of $M\times N$ complex and real matrices.
$\vec{1}_{M\times N}$ represents an $M\times N$ matrix with all entries equal to 1.
$j=\sqrt{-1}$ denotes the imaginary unit of complex numbers.
The distribution of a circularly symmetric complex Gaussian (CSCG) random variable with mean 0 and variance $\sigma^2$ is denoted by $N_{\mathbb{C}}(0,\sigma^2)$, and $N(0,\sigma^2)$ denotes the real-valued Gaussian distribution.
$\text{Unif}(a,b)$ denotes the continuous uniform distribution in the interval $[a,b]$.
${\rm {round}}(\cdot)$ represents rounding to the nearest integer.
$\|\cdot\|_0$ denotes the $l_0$ norm.
 $\lceil\cdot\rceil$ and $\lfloor\cdot\rfloor$ denote the ceiling and floor operations, respectively.
 $\otimes$ denotes the Kronecker product operation.
  $\text{vec}(\cdot)$ and $\text{diag}(\cdot)$ denote the vectorization operation and diagonalization operation, respectively.

\section{System Model for RAA-based Wireless Communication}
\begin{figure*}[htbp]
\centering
\subfigure[]{
\includegraphics[height=3.2cm]{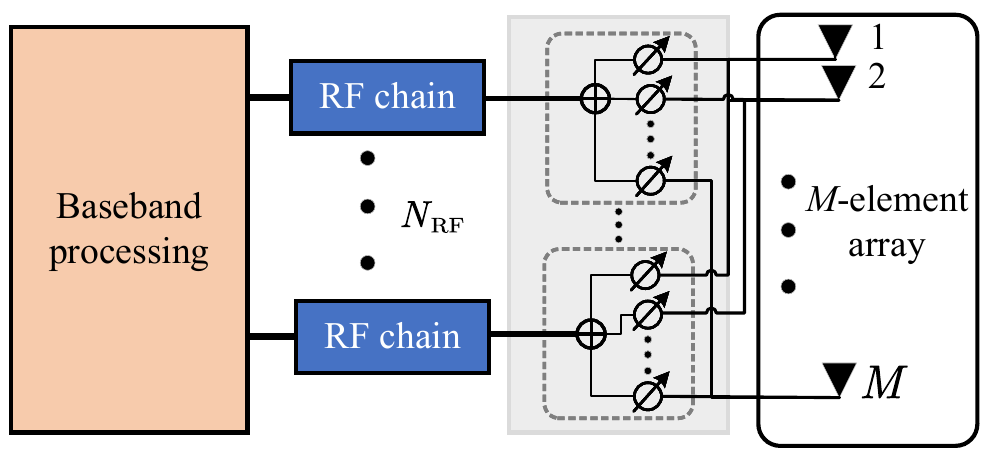}}
\hspace{0.3cm}
\subfigure[]{
\includegraphics[height=3.2cm]{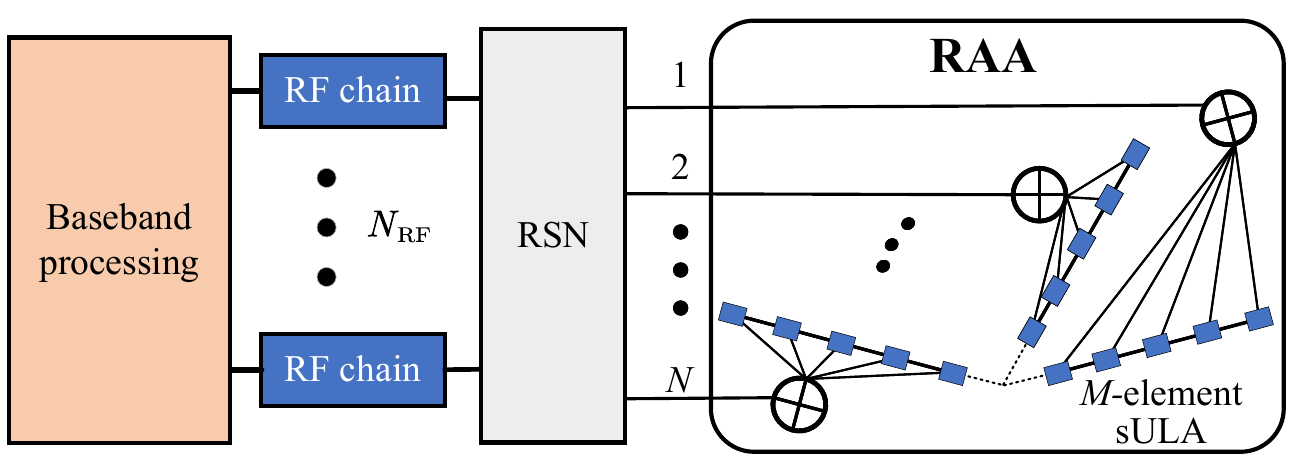}}
	\caption{MIMO architectures with $N_{\text{RF}}$ chains. (a) The conventional fully-connected HBF architecture based on phase shifters. (b) The proposed RAA architecture without any phase shifters.}
	\label{Architecture}
\end{figure*}

As shown in Fig. \ref{Architecture}, we consider a wireless communication system with $N_{\text{RF}}$ RF chains supporting an $M$-element antenna array, with $N_{\text{RF}}\leq M$.
The HBF is a classic architecture to meet such a requirement by combining low-dimensional baseband digital beamforming with high-dimensional analog beamforming.
The extensively studied fully-connected HBF architecture based on phase shifters is shown in Fig. \ref{Architecture}(a), in which each antenna element is connected to $N_{\text{RF}}$ phase shifters.
As a result, this architecture requires a total of $N_{\text{RF}}\times M$ phase shifters, which is huge for large-array systems. For example, in a system with $M=128$ antennas and $N_{\text{RF}}=16$ RF chains, a total of 2048 phase shifters are required, resulting in prohibitive hardware costs and power consumption. 

To reduce hardware costs while enhancing the wireless performance, we propose a novel RAA-based wireless communication system, as shown in Fig. \ref{Architecture}(b). The proposed RAA is composed of $N$ so-called sULAs, each comprising $M$ antenna elements. The $N$ sULAs are deliberately arranged in a ray-like structure with carefully designed ray orientations, as will be detailed in Section III. Within each sULA, all the $M$ antenna elements are directly connected without relying on any analog or digital processing, so as to justify the term sULA. Additionally, a RSN is introduced to dynamically select $N_{\text{RF}}$ sULAs to connect to the RF chains, thereby enabling subsequent baseband processing. The RSN is characterized by a selection matrix  $\vec{S}\in\{0,1\}^{N_{\text{RF}}\times N}$,
which satisfies that $\left\|[\vec{S}]_{i,:}\right\|_0=1$ and $\left\|[\vec{S}]_{:,n}\right\|_0\leq1$
, $1\leq i\leq N_{\text{RF}}$ and $1\leq n\leq N$.


\begin{figure}[htbp]
\centering
\includegraphics[width=0.9\linewidth]{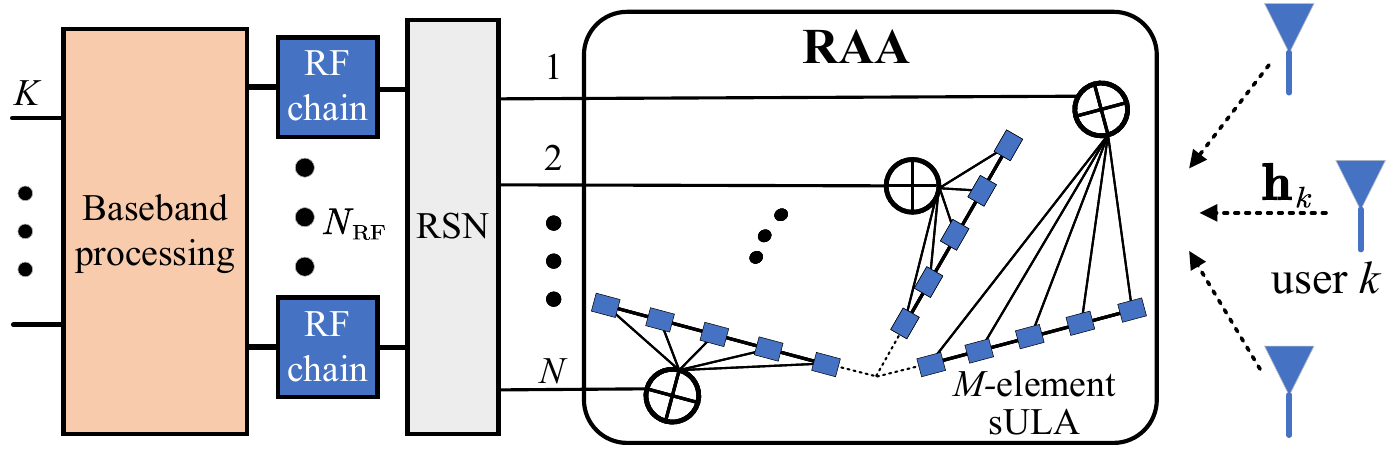}
	\caption{Uplink multi-user communications based on the proposed RAA architecture.}
	\label{Kuser}
\end{figure}

The uplink multi-user communications based on the proposed RAA architecture is first considered, as shown in Fig. \ref{Kuser}, where a base station (BS) equipped with the proposed RAA serves $K$ single-antenna users.
Before undergoing the RSN and baseband processing, the received signal vector at the $N$ ports, denoted by $\vec{x}\in\mathbb{C}^{N\times 1}$, is expressed as
\begin{equation}\label{Signal}
    \vec{x}=\sum\nolimits_{k=1}\nolimits^{K}\vec{h}_{k}s_k+\vec{z},
\end{equation}
where $\vec{h}_k\in\mathbb{C}^{N \times 1}$ represents the effective channel between user $k$ and the $N$ sULA ports, which will be modeled in Section III;
$s_k$ denotes the  independent and identically distributed (i.i.d.) information-bearing symbol of user $k$, satisfying $\mathbb{E}[\vec{s}\vec{s}^H]=P_t\vec{I}_K$ with $P_t$ representing the transmit power of user $k$ and $\vec{s}=[s_k]_{1\leq k\leq K}\in \mathbb{C}^{K\times1}$;
and $\vec{z}\in \mathbb{C}^{N\times 1}\sim N_{\mathbb{C}}(0,\sigma_0^2\vec{I}_{N})$ is the effective noise vector at the $N$ sULAs, following i.i.d. CSCG distribution with variance $\sigma_0^2$.
After applying the ray selection and baseband processing, the resulting signal of user $k$ at the BS can be written as
\begin{equation}\label{channel modelX}
    y_k=\vec{w}_k^H \vec{S}\vec{h}_ks_k+\vec{w}_k^H \vec{S}\sum\nolimits_{i\neq k}\vec{h}_is_i
    +\vec{w}_k^H \vec{S}\vec{z},
\end{equation}
where $\vec{w}_k\in \mathbb{C}^{N_{\text{RF}} \times 1}$ is the baseband beamforming for user $k$.

Compared to the conventional HBF architecture, the proposed RAA system in (\ref{channel modelX}) has two important differences.
First, the channel vector $\vec{h}_k$ in the RAA architecture refers to the effective channel between user $k$ and the  $N$ sULAs, in contrast to the physical channel between user $k$ and the $M$-element array in the conventional HBF.
Second, the signal processing matrix $\vec{S}$ in the RF domain for the RAA architecture is a binary matrix, whereas the analog beamforming matrix in the HBF architecture consists of complex-valued beamforming coefficients with constant modulus since they correspond to the phase shifting introduced by the phase shifters.

According to (\ref{channel modelX}), the SINR of user $k$ for the RAA-based communications is
\begin{equation}\label{channel model3}
     \text{SINR}_k=\frac{\bar{P}'_t\left|\vec{w}_k^H \vec{S}\vec{h}_k
    \right|^2}{\bar{P}'_t\sum_{i\neq k}\left|\vec{w}_k^H \vec{S}\vec{h}_i
    \right|^2+\|\vec{w}_k^H \vec{S}\|^2},
\end{equation}
where $\bar{P}'_t=\frac{P_t}{\sigma_0^2}$.
For the special case of single-user communication, the receive SNR is given by
 \begin{equation}\label{SNR}
\text{SNR}=\frac{\bar{P}'_t\left|\vec{w}^H\vec{S}\vec{h}\right|^2}{\vec{w}^H \vec{S}\vec{S}^H \vec{w}}= \frac{\bar{P}'_t\left|\vec{w}^H\vec{S}\vec{h}\right|^2}{\vec{w}^H  \vec{w}},
 \end{equation}
where the second equality holds due to $[\vec{S}]_{i,:}[\vec{S}]_{i,:}^H=1$ and $[\vec{S}]_{i,:}[\vec{S}]_{j,:}^H=0$, $\forall j\neq i$, thus $\vec{S}\vec{S}^H=\vec{I}_{N_{\text{RF}}}$.

In the following, we first present the specific design of the RAA architecture. Based on this design, we then provide the modeling of the effective channel $\vec{h}_k$ in (\ref{channel modelX}) and discuss the design of joint RAA beamforming and ray selection.

\section{Proposed RAY ANTENNA ARRAY}
\subsection{RAA Basic Structure}
\begin{figure}[h]
\centering
\includegraphics[width=0.8\linewidth]{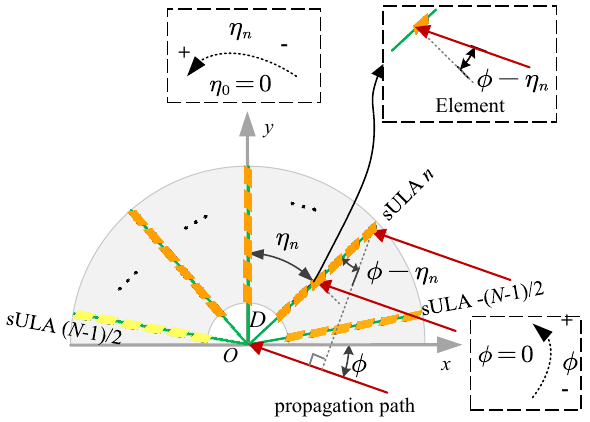}
	\caption{The proposed RAA architecture is composed of $MN$ antenna elements, which are arranged into $N$ rays. Each ray corresponds to one $M$-element sULA with all antenna elements directly connected. }
	\label{2Darray}
\end{figure}
As shown in Fig. \ref{2Darray}, the proposed RAA is composed of $NM$ antenna elements arranged into $N$ sULAs. Each sULA comprises $M$ antenna elements, with adjacent elements spaced by $d=\frac{\lambda}{2}$, where $\lambda$ denotes the signal wavelength.
The $N$ sULAs are arranged in a ray-like configuration, where each sULA is oriented along a carefully designed direction.
Without loss of generality, a Cartesian coordinate system is established, where the $N$ sULAs are placed symmetrically around the $y$-axis.
For simplicity of notation, we assume that $N$ is an odd number.
Denote $\mathcal{N}=\left\{-\frac{N-1}{2},\cdots,0,\cdots,\frac{N-1}{2}\right\}$ as the index set of $N$ sULAs.
The orientation of sULA $n$ w.r.t. the positive $y$-axis is denoted by $\eta_n\in\left[-\eta_{\text{max}},\eta_{\text{max}}\right]$, 
$\forall n\in\mathcal{N}$.
By letting sULA $n=0$ align with the positive $y$-axis and serve as the reference direction, we have $\eta_{0}=0$.
Note that the orientation $\eta_n$ takes negative values in the first and fourth quadrants, while positive values in the second and third quadrants.
To ensure a minimum physical separation between antenna elements across sULAs,
the first element of each sULA is placed at a distance $D$ from the origin.
  Accordingly, the proposed RAA architecture can be fully characterized by the parameter set $\left\{M,N,D,\{\eta_n\}_{n\in\mathcal{N}},G(\cdot)\right\}$, where $G(\cdot)$ denotes the radiation pattern of an individual antenna element in the proposed RAA.

Let $\phi\in\left[-\phi_{\text{max}},\phi_{\text{max}}\right]$ denote the AoA/AoD of a propagation path w.r.t the positive $x$-axis, with $\phi_{\text{max}}<0.5\pi$. Note that for notational convenience, the sULA orientation angle $\eta_n$ and the path angle $\phi$ are purposely defined based on different reference directions. Additionally, the maximum ray orientation is set equal to the maximum path angle, i.e., $\eta_{\text{max}}=\phi_{\text{max}}$.
Therefore, the array response vector of sULA $n$ for the path angle $\phi$, denoted by $\vec{a}(\phi,\eta_n) \in\mathbb{C}^{M\times 1}$, is given by
\begin{equation}\label{steering}
\vec{a}(\phi,\eta_n)=\bigl[1,e^{j\pi\sin(\phi-\eta_n)},\cdots,e^{j\pi(M-1)\sin(\phi-\eta_n)}\bigl]^T,
\end{equation}
where $\phi-\eta_n$ is the path angle w.r.t. the $n$th sULA, as illustrated in Fig. \ref{2Darray}. Thus, the array response matrix for the path angle $\phi$, denoted by $\vec{A}(\phi)\in\mathbb{C}^{M\times N}$, is given by
\begin{equation}\label{steering_A}
\vec{A}(\phi)=\left[\vec{a}(\phi,\eta_n)\right]_{n\in\mathcal{N}}\times\text{diag}\left(\vec{b}(\phi)\right),
\end{equation}
where $\vec{b}(\phi)=\left[b(\phi-\eta_n)\right]_{n\in\mathcal{N}}\in\mathbb{C}^{N\times 1}$ captures the response of the reference element within each sULA, say the first element, given by
\begin{equation}\label{steering_bn}
b(\phi-\eta_n)=e^{j\frac{2\pi}{\lambda}D\sin(\phi-\eta_n)}\sqrt{G(\phi-\eta_n)},
\end{equation}
with the component $e^{j\frac{2\pi}{\lambda}D\sin(\phi-\eta_n)}$ corresponding to the phase shifting of
the first antenna element w.r.t. the origin, and $G(\zeta)$ denoting the antenna element pattern, which 
is characterized by the peak antenna gain $G(0)$, the 3dB beamwidth $\phi_{\text{3dB}}$, and the total power gain along azimuth or elevation direction $G_{\text{sum}}=\int_{-\zeta_{\text{max}}}^{\zeta_{\text{max}}}G(\zeta)d\zeta$, with $\zeta_{\text{max}}=\pi$ for the azimuth dimension and $\zeta_{\text{max}}=0.5\pi$ for the elevation dimension.

In the proposed RAA architecture,
all the $M$ antenna elements within each sULA are directly connected.
Thus, the resulting outputs for the $N$ sULAs are denoted by $\vec{f}(\phi)=[f(\phi,\eta_n)]_{n\in\mathcal{N}}\in\mathbb{C}^{N\times 1} $, with $f(\phi,\eta_n)$ representing the output of the $n$th sULA and expressed as
\begin{equation}\label{steering_eff00}
\begin{aligned}
f(\phi,\eta_n)&=\vec{1}^T_{M\times 1}\vec{a}(\phi,\eta_n)b(\phi-\eta_n)\\
&=\Bigl(\sum\nolimits_{m=0}\nolimits^{M-1}e^{j\pi m\sin(\phi-\eta_n)}\Bigl)b(\phi-\eta_n)\\
&=\underbrace{MH_M(\sin(\phi-\eta_n))}_{r(\phi,\eta_n)}b(\phi-\eta_n),
\end{aligned}
\end{equation}
where $H_M\left(x\right)=\frac{1}{M}\sum_{m=0}^{M-1}e^{j\pi mx}$ is the Dirichlet kernel function, given by
\begin{equation}\label{H_M}
\begin{aligned}
H_M\left(x\right)=e^{j\frac{\pi}{2}(M-1)x}\frac{\sin\left(\frac{\pi}{2}Mx\right)}{M\sin\left(\frac{\pi}{2}x\right)}.
\end{aligned}
\end{equation}

\begin{figure}[htbp]
\centering
\includegraphics[width=0.7\linewidth]{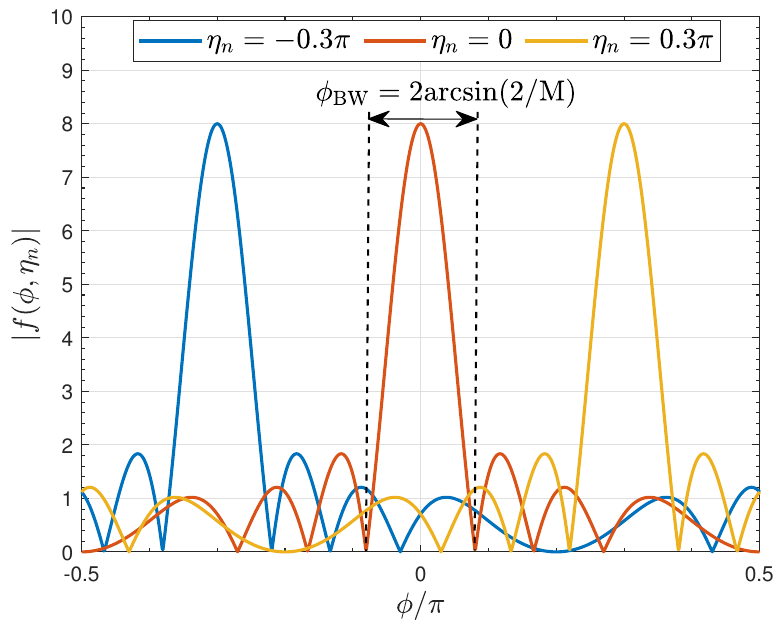}
	\caption{Beam patterns $|f(\phi,\eta_n)|$ in (\ref{steering_eff00}) for $\eta_n=-0.3\pi$, $0$, and $0.3\pi$, where $M=8$ and $G(\zeta)=1$, $\forall \zeta$.}
	\label{2DRAA_1}
\end{figure}

The beam patterns $|f(\phi,\eta_n)|$ in (\ref{steering_eff00}) for $\eta_n=-0.3\pi$, $0$, and $0.3\pi$, are illustrated in Fig. \ref{2DRAA_1}. 
It is observed that the proposed RAA can form a beam towards the direction aligned with the physical orientation of the sULA $\eta_n$, without relying on any analog or digital beamforming.


\subsection{RAA Parameter Design}
In this section, we aim to design the parameters of the proposed RAA architecture.
For any given sULA $n$ with orientation $\eta_n$, the path angle $\phi$ that has the maximum and minimum responses  $|r(\phi,\eta_n)|$ in (\ref{steering_eff00}) can be obtained as follows.

\begin{lem}\label{lemma01}
For any given sULA $n$ with orientation $\eta_n$,
\begin{equation}\label{hu}
\begin{aligned}
  &\underset{|\phi|\leq\phi_{\text{max}}}{\text{max}}(|r(\phi,\eta_n)|)=M, \text{ and the maximum} \\
  &\text{is achieved if and only if } \phi=\eta_n.
  \end{aligned}
\end{equation}
\end{lem}
\begin{pproof}
The
 function $|H_M(x)|$ in (\ref{H_M}) achieves the maximum value 1 when $x=0,\pm2,\pm4,\cdots$.
Consequently, $|H_M(\sin(\phi-\eta_n))|$  in (\ref{steering_eff00}) achieves the maximum value 1 if and only if $\sin(\phi-\eta_n)=0$, since $\sin(\phi-\eta_n)\in[-1,1]$. Thus, $|r(\phi,\eta_n)|$ in (\ref{steering_eff00}) achieves the maximum value $M$ if and only if $\phi=\eta_n$ due to $|\phi-\eta_n|<\pi$ with $\eta_{\max}=\phi_{\max}<0.5\pi$.
    $\hfill\blacksquare$
\end{pproof}

\begin{lem}\label{lemma02}
For any given sULA $n$ with orientation $\eta_n$,
\begin{equation}\label{hu}
\begin{aligned}
  &\underset{|\phi|\leq\phi_{\text{max}}}{\text{min}}(|r(\phi,\eta_n)|)=0, \text{ and the minimum is achieved} \\
  &\text{if and only if } \phi=\arcsin\Bigl(\frac{2p}{M}\Bigl)+\eta_n, \enspace \forall p\in\mathcal{P}(\eta_n),
  \end{aligned}
\end{equation}
where $\mathcal{P}(\eta_n)=\bigl\{\bigl\lceil\frac{M\cdot\chi_{\text{min}}(\eta_n)}{2}\bigl\rceil,\cdots,-1,1,\cdots, \bigl\lfloor\frac{M\cdot\chi_{\text{max}}(\eta_n)}{2}\bigl\rfloor\bigl\}$ with
$\chi_{\text{min}}(\eta_n)=\underset{|\omega|\leq\phi_{\text{max}}}{\text{min}}\sin(\omega-\eta_n)$
and $\chi_{\text{max}}(\eta_n)=\underset{|\omega|\leq\phi_{\text{max}}}{\text{max}}\sin(\omega-\eta_n)$.

\end{lem}
\begin{pproof}
The function $|H_M(x)|$ in (\ref{H_M}) achieves the minimum value 0 if and only if $x=\frac{2p}{M}$ with  $p=\pm1,\pm2,\cdots$. Thus,
$|r(\phi,\eta_n)|$ in (\ref{steering_eff00}) achieves the minimum value 0 if and only if $\phi=\arcsin\left(\frac{2p}{M}\right)+\eta_n$, $\forall p\in\mathcal{P}(\eta_n)$.
    $\hfill\blacksquare$
\end{pproof}

Thus, the null-to-null mainlobe beamwidth of sULA $n$ in the proposed RAA can be obtained by taking $p=\pm 1$ in Lemma \ref{lemma02}, which is given below.

\begin{cor}\label{lemma2DRAA0}
For given any sULA $n$ with orientation $\eta_n$,
the null-to-null mainlobe beamwidth is
\begin{equation}\label{BW}
\phi_{\text{BW}}=2\arcsin\Bigl(\frac{2}{M}\Bigl).
\end{equation}
\end{cor}

\begin{figure}[htbp]
\centering
\includegraphics[width=0.7\linewidth]{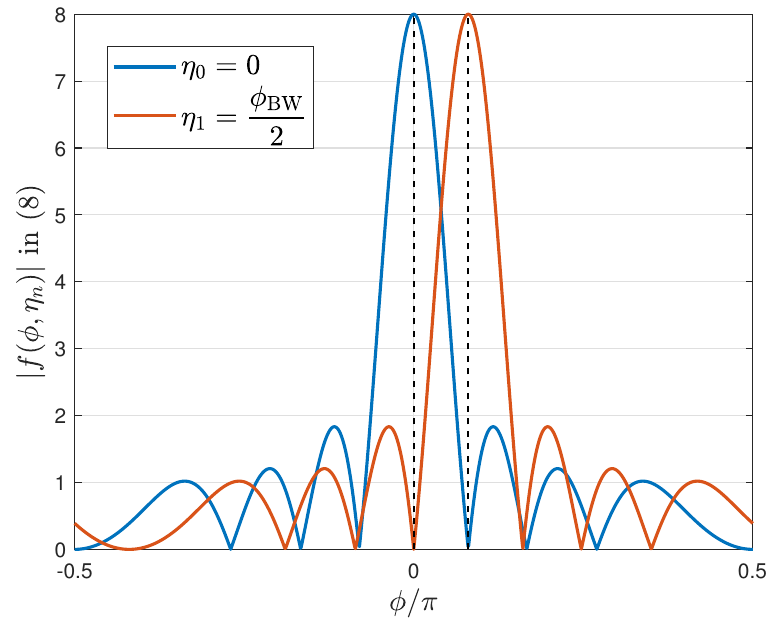}
	\caption{Orientation design principle for adjacent sULAs in RAA, where $M=8$ and $G(\zeta)=1$, $\forall \zeta$.}
	\label{2DRAA_criterion}
\end{figure}

It can be seen from Corollary \ref{lemma2DRAA0} that all sULAs have identical mainlobe beamwidths, as illustrated in Fig. \ref{2DRAA_1}.
Based on this, we aim to design the ray orientations $\eta_n$. Due to the nonlinear nature of the function $\arcsin(\cdot)$, the null points of $|r(\phi,\eta_n)|$ are not evenly spaced, i.e., $\arcsin\left(\frac{2p}{M}\right)+\eta_n\neq p\times\arcsin\left(\frac{2}{M}\right)+\eta_n$, $\forall p\in\mathcal{P}(\eta_n)$.
Consequently, it is impossible to perfectly align the null points of sULA $n$ with those of all other sULAs.
Therefore, to ensure orthogonality between adjacent sULAs, say sULA $n$ and sULA $n-1$, the orientation $\eta_{n}$ for given $\eta_{n-1}$ is designed so that the first null point of sULA $n$ aligns with the peak of sULA $n-1$, as  illustrated in Fig. \ref{2DRAA_criterion}. As a result, we have $|\eta_n-\eta_{n-1}|=0.5\phi_{\text{BW}}=\arcsin\left(\frac{2}{M}\right)$.

Thus, the total number of sULAs required to cover the total angle range of $2\phi_{\text{max}}=2\eta_{\max}$ can be determined
\begin{equation}\label{N}
N=2 \Bigl\lfloor\frac{\eta_{\text{max}}}{\arcsin(2/M)}\Bigl\rfloor+1,
\end{equation}
and the orientation of sULA $n$ is designed to
\begin{equation}\label{eta_n}
\eta_n=n\times\arcsin\Bigl(\frac{2}{M} \Bigl), \enspace \forall n\in\mathcal{N}.
\end{equation}


In addition, to ensure that the distance between the first antenna elements of adjacent sULAs is no smaller than half wavelength, the distance $D$ needs to satisfy
\begin{equation}\label{distanceD}
D\geq \frac{\lambda}{4\sin(0.5\arcsin\left(\frac{2}{M}\right))}. 
\end{equation}

When $M\gg 1$, we have
\begin{equation}
\begin{aligned}
&N\approx \lfloor M\eta_{\text{max}}\rfloor,\enspace \eta_n\approx\frac{2n}{M},\text{ and }D\approx\frac{M\lambda}{4}, \enspace \forall n\in\mathcal{N}.
\end{aligned}
\end{equation}

So far, the parameters $\left(N,D,\{\eta_n\}_{n\in\mathcal{N}}\right)$ of the proposed RAA can be determined based on
(\ref{N})-(\ref{distanceD}). 
The remaining task is to determine the antenna element pattern $G(\zeta)$.
It is worth noting that in the conventional antenna architecture such as ULA, since the antenna array is shared by all signal directions,
omnidirectional or at least wide-coverage antenna elements are required.
In contrast, the proposed RAA consists of multiple sULAs with different orientations, where the antenna gain  $G(\phi-\eta_n)$ in (\ref{steering_bn}) also involves the sULA orientation $\eta_n$. This structural feature naturally decomposes the total angular coverage range $[-\phi_{\max}, \phi_{\max}]$ among the $N$ sULAs.
As a result, the RAA architecture offers additional DoF in designing the antenna element pattern $G(\zeta)$. 
Intuitively, compared to the conventional antenna arrays, the antenna elements of the proposed RAA can have higher directivity, since they are only responsible for a small portion of the total angle coverage range. 
This will be rigorously demonstrated in the following.

The objective of designing $G(\zeta)$ is to ensure sufficient overall beamforming gain over the entire angle coverage range.
Based on (\ref{steering_eff00}), the maximum beamforming gain for signal direction $\phi$ is defined as
\begin{equation}\label{we}
f(\phi)=\underset{\forall n\in\mathcal{N}}{\text{max}}\enspace \left|f(\phi,\eta_n)\right|, \enspace \forall \phi\in[-\phi_{\text{max}},\phi_{\text{max}}],
\end{equation}
 which corresponds to the overall beamforming gain including the antenna element pattern by selecting the best sULA.
 The design criterion for $G(\zeta)$ is then to guarantee that this effective gain is no lower than a predetermined threshold $\varepsilon_0$ over the entire angle range $[-\phi_{\text{max}},\phi_{\text{max}}]$, i.e.,
\begin{equation}\label{we01}
\begin{aligned}
\underset{-\phi_{\text{max}}\leq\phi\leq\phi_{\text{max}}}{\text{min}}f(\phi)\geq \varepsilon_0.
\end{aligned}
\end{equation}

To satisfy (\ref{we01}), the antenna element pattern of RAA $G(\zeta)$ needs to be carefully designed, as summarized in Theorem \ref{lemmaX0001}.

\begin{thm}\label{lemmaX0001}
Under the following two mild assumptions for $G(\zeta)$: (i) $G(\zeta)=G(-\zeta)$, and $G(\zeta)$ is monotonically non-increasing with $|\zeta|$ for $|\zeta|\leq 0.5\phi_{\text{3dB}}$, (ii) only two half-power ($-3$ dB) points exist, i.e., $\left|\{\zeta:G(\zeta)=0.5G(0)\}\right|=2$, a sufficient condition to ensure (\ref{we01}) is
\begin{equation}\label{2DRAA_AP}
\begin{aligned}
 &G(0.5\phi_{\text{3dB}}) \geq \Bigl(\frac{\varepsilon}{\left|H_M\left(\sin\left(0.5\arcsin\left(\frac{2}{M}\right)\right)\right)\right|}\Bigl)^2, \\
 &\text{and } \phi_{\text{3dB}}\geq\arcsin\Bigl(\frac{2}{M}\Bigl),
\end{aligned}
\end{equation}
where $\varepsilon=\varepsilon_0/M$.
\end{thm}
\begin{pproof}
    Please refer to Appendix \ref{AA}.
    $\hfill\blacksquare$
\end{pproof}


One commonly used model for $G(\zeta)$ is the 3GPP antenna element pattern \cite{3GPPchannel}, which is modeled in dB as
\begin{equation}\label{G}
G_{\text{dB}}(\zeta)=G_{\text{dB}}(0)-\text{min}\bigl\{12\left(\zeta/\phi_{\text{3dB}}\right)^2,A_{\text{max}}\bigl\},
\end{equation}
with $G_{\text{dB}}(0)=10\log_{10}(G(0))$ and $A_{\text{max}}=$30 dB describing the peak gain in dB and front-to-back attenuation, respectively.
It is not difficult to verify that $G(\zeta)$ in (\ref{G}) satisfy assumptions (i) and (ii).
Then based on the law of power conservation with $G_{\text{sum}}$, the peak gain $G(0)$ can be expressed in terms of $\phi_{\text{3dB}}$ and $G_{\text{sum}}$, given in Theorem \ref{AntennaPattern}.


\begin{thm}\label{AntennaPattern}
Given the 3dB beamwidth $\phi_{\text{3dB}}$, total power gain $G_{\text{sum}}$, and antenna element pattern in (\ref{G}), the peak gain $G(0)$ is expressed as
\begin{equation}\label{G00}
G(0)\approx\frac{G_{\text{sum}}}{1.066\phi_{\text{3dB}}}. 
\end{equation}
\end{thm}
\begin{pproof}
    Please refer to Appendix \ref{2AP1}.
    $\hfill\blacksquare$
\end{pproof}

Equation (\ref{G00}) reveals an inverse relationship between the 3dB beamwidth $\phi_{\text{3dB}}$ and peak gain $G(0)$, indicating that narrower beam leads to higher peak gain, as expected. 
Furthermore, the second equation in (\ref{2DRAA_AP}) shows that for the proposed RAA, increasing the number of antennas $M$ for each sULA allows for a reduction in $\phi_{\text{3dB}}$, since each sULA only needs to cover a smaller angular range.

When $M\gg1$, (\ref{2DRAA_AP}) can be rewritten as
\begin{equation}\label{2DRAA_0AP}
\begin{aligned}
 G(0.5\phi_{\text{3dB}}) \geq \Bigl(\frac{\varepsilon}{\left|H_M\left(1/M\right)\right|}\Bigl)^2  \text{ and } \phi_{\text{3dB}}\geq\frac{2}{M}.
\end{aligned}
\end{equation}

\begin{figure}[htbp]
\centering
\includegraphics[width=0.6\linewidth]{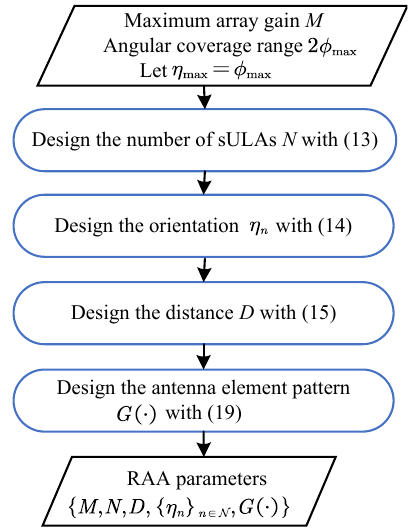}
	\caption{The overall procedure of RAA parameter design. }
	\label{overall}
\end{figure}

The complete procedure for designing the parameters $\left\{M,N,D,\{\eta_n\}_{n\in\mathcal{N}},G(\cdot)\right\}$ of the proposed RAA architecture is illustrated in Fig. \ref{overall}.

\subsection{Comparison with the Conventional ULA}
Next, we compare the proposed RAA with the conventional ULA with HBF, where the analog beamforming is implemented using discrete Fourier transform (DFT) codebook. 
The array response vector of the $M$-element ULA for the path angle $\phi$, denoted by $\vec{a}_{\text{ULA}}(\phi)\in \mathbb{C}^{M\times 1}$, is given by
\begin{equation}\label{ULA}
  \vec{a}_{\text{ULA}}(\phi)=\bigl[1,e^{j\pi\sin\phi},\cdots,e^{j\pi(M-1)\sin\phi}\bigl]^T.
\end{equation}

 The DFT codebook is denoted by $\vec{C}_{\text{DFT}}=[\vec{c}(\phi_n)]_{n\in \mathcal{N}'}\in \mathbb{C}^{M\times N'}$, where $\vec{c}(\phi_n)\in \mathbb{C}^{M\times 1}$ is the $n$th DFT codeword, given by
 \begin{equation}\label{DFT}
   \vec{c}(\phi_n)=\bigl[1,e^{j\pi\sin\phi_n},\cdots,e^{j\pi(M-1)\sin\phi_n}\bigl]^T, \forall n\in \mathcal{N}',
 \end{equation}
 with $\sin\phi_n=\frac{2n}{M}$, $N'=2\bigl\lfloor\frac{\sin\phi_{\text{max}}}{2/M} \bigl\rfloor+1$ being the number of DFT codewords \cite{pillai2012array}, and $\mathcal{N}'=\left\{-\frac{N'-1}{2},\cdots,0,\cdots,\frac{N'-1}{2}\right\}$ being the index set of DFT codewords.
 Consequently, considering the impact of antenna element pattern, the overall beam pattern of the ULA for DFT codeword $n$, denoted by $f_{\text{ULA}}(\phi,\phi_n)$, can be expressed as
\begin{equation}\label{steeringV1X}
\begin{aligned}
f_{\text{ULA}}(\phi,\phi_n)&=\sqrt{G_{\text{ULA}}(\phi)}\vec{c}^H(\phi_n)\vec{a}_{\text{ULA}}(\phi)\\
&=\sqrt{G_{\text{ULA}}(\phi)}\underbrace{MH_M(\sin\phi-\sin\phi_n)}_{r_{\text{ULA}}(\phi,\phi_n)} ,
\end{aligned}
\end{equation}
where $G_{\text{ULA}}(\phi)$ is the antenna element pattern in the ULA, characterized by the peak antenna gain $G_{\text{ULA}}(0)$, the 3dB beamwidth $\phi_{\text{3dB}}^{\text{ULA}}$, and the total power gain  $G^{\text{ULA}}_{\text{sum}}$. 


For any given DFT codeword $n$, the path angle $\phi$ that has the maximum and minimum responses  $|r_{\text{ULA}}(\phi,\phi_n)|$ in (\ref{steeringV1X}) can be obtained as follows.

\begin{lem}\label{lemma001}
For any given DFT codeword $n$,
\begin{equation}\label{hu}
\begin{aligned}
  &\underset{|\phi|\leq\phi_{\text{max}}}{\text{max }}|r_{\text{ULA}}(\phi,\phi_n)|=M, \text{ and the maximum  } \\
  &\text{is achieved if and only if }\phi=\phi_n.
  \end{aligned}
\end{equation}
\end{lem}
\begin{pproof}
The
 function $|H_M(x)|$ in (\ref{H_M}) achieves the maximum value 1 when $x=0,\pm2,\pm4,\cdots$.
Consequently, $|H_M(\sin\phi-\sin\phi_n)|$ in (\ref{steeringV1X}) achieves the maximum value 1 if and only if $\phi=\phi_n$ due to $\phi_{\text{max}}<0.5\pi$.
    $\hfill\blacksquare$
\end{pproof}

\begin{lem}\label{lemma002}
For any given DFT codeword $n$,
\begin{equation}\label{hu}
\begin{aligned}
  &\underset{|\phi|\leq\phi_{\text{max}}}{\text{min }}|r_{\text{ULA}}(\phi,\phi_n)|=0, \text{and the minimum is achieved} \\
  &\text{if and only if } \phi=\arcsin\Bigl(\sin\phi_n+\frac{2p}{M}\Bigl), \forall p\in\mathcal{P}(\phi_n),
  \end{aligned}
\end{equation}
where $\mathcal{P}(\phi_n)=\bigl\{\left\lceil -\frac{M}{2} (\sin\phi_{\text{max}}+\sin\phi_n)\right\rceil,\cdots,-1,1,$
$\cdots,\left\lfloor \frac{M}{2} (\sin\phi_{\text{max}}-\sin\phi_n)\right\rfloor\bigl\}$.
\end{lem}
\begin{pproof}
If and only if $\sin\phi-\sin\phi_n=\frac{2p}{M}$, $\forall p\in \mathcal{P}(\phi_n)$, $|H_M(\sin\phi-\sin\phi_n)|$ in (\ref{steeringV1X}) can achieve the minimum value 0. Thus,
if and only if $\phi=\arcsin\left(\sin\phi_n+\frac{2p}{M}\right)$, $|r_{\text{ULA}}(\phi,\phi_n)|$ in (\ref{steeringV1X}) achieves the minimum value 0, $\forall p\in \mathcal{P}(\phi_n)$.
    $\hfill\blacksquare$
\end{pproof}

Thus, the null-to-null mainlobe beamwidth of the conventional ULA with HBF can be obtained by taking $p=\pm1$ in Lemma \ref{lemma002}, given in Corollary \ref{ULABW}.
\begin{cor}\label{ULABW}
For any given DFT codeword $n$, the null-to-null mainlobe beamwidth of the conventional ULA is
\begin{equation}\label{BWULA}
\begin{aligned}
\phi^{\text{ULA}}_{\text{BW},n}&=\arcsin\Bigl(\sin\phi_n+\frac{2}{M}\Bigl)-\arcsin\Bigl(\sin\phi_n-\frac{2}{M}\Bigl)\\
&\overset{(a)}{\approx}2\arcsin\Bigl(\frac{2}{M}\Bigl)+\frac{2/M}{(1-(2/M)^2)^{\frac{3}{2}}}\phi^2_n,
\end{aligned}
\end{equation}
where ``$\overset{(a)}{\approx}$'' holds from the second-order Taylor approximation.
It is found from (\ref{BWULA}) that $\phi^{\text{ULA}}_{\text{BW},n}\geq 2\arcsin\left(\frac{2}{M}\right)$, where ``='' holds if and only if  $\phi_n=0$.
In addition, the beamwidth $\phi^{\text{ULA}}_{\text{BW},n}$ increases monotonically with $|\phi_n|$, resulting in poor angular resolution for edge users.
\end{cor}


To facilitate a fair comparison between the classic ULA and the proposed RAA, we also analyze the antenna element pattern in the conventional ULA. Similarly, we first define the maximum beamforming gain in the conventional ULA for the signal direction $\phi$ as follows
\begin{equation}\label{DFTcoverage}
f_{\text{ULA}}(\phi)=\underset{\forall n\in\mathcal{N}'}{\text{max}}\enspace \left|f_{\text{ULA}}(\phi,\phi_n)\right|, \forall \phi\in[-\phi_{\text{max}},\phi_{\text{max}}].
\end{equation}

 The antenna element pattern in the ULA $G_{\text{ULA}}(\zeta)$ also needs to guarantee that $f_{\text{ULA}}(\phi)$ is no lower than the threshold $\varepsilon_0$ over the entire angle range, i.e.,
\begin{equation}\label{DFTcoverage1}
\begin{aligned}
\underset{-\phi_{\text{max}}\leq\phi\leq\phi_{\text{max}}}{\text{min}}f_{\text{ULA}}(\phi)\geq \varepsilon_0.
\end{aligned}
\end{equation}


To satisfy (\ref{DFTcoverage1}), the antenna element pattern of the ULA $G_{\text{ULA}}(\zeta)$ needs to be carefully designed, as summarized in Theorem \ref{lemmaX002}.


\begin{thm}\label{lemmaX002}
Under the following mild assumption 
(iii)
$G_{\text{ULA}}(\zeta)=G_{\text{ULA}}(-\zeta)$, and $G_{\text{ULA}}(\zeta)$ is monotonically non-increasing with $|\zeta|$ for $|\zeta|\leq 0.5\phi_{\text{3dB}}^{\text{ULA}}$, a sufficient condition to ensure (\ref{DFTcoverage1}) is given as follows, with $\varepsilon_0=M\varepsilon$
\begin{equation}\label{ULA_AP}
\begin{aligned}
G_{\text{ULA}}(0.5\phi_{\text{3dB}}^{\text{ULA}}) \geq \Bigl(\frac{\varepsilon}{\left|H_M(1/M)\right|}\Bigl)^2 \text{ and } \phi_{\text{3dB}}^{\text{ULA}}\geq2\phi_{\text{max}}.
\end{aligned}
\end{equation}
\end{thm}
\begin{pproof}
    Please refer to Appendix \ref{2DLE1}.
    $\hfill\blacksquare$
\end{pproof}

It is found from (\ref{ULA_AP}) that the conventional ULA requires wide-coverage antenna elements, since it needs to cover the total angular coverage range $2\phi_{\text{max}}$.

Compared to the conventional ULA, 
the proposed RAA exhibits following three significant advantages. 

\begin{itemize}
    \item \emph{Finer and uniform angular resolution}: By comparing the beamwidth of the proposed RAA $\phi_{\text{BW}}$ in (\ref{BW}) to that of the ULA $\phi^{\text{ULA}}_{\text{BW},n}$ in (\ref{BWULA}), we have
        \begin{equation}\label{rt}
         \phi^{\text{ULA}}_{\text{BW},n}\geq \phi_{\text{BW}},\enspace \forall n\in\mathcal{N}',
         \nonumber
        \end{equation}
        where the equality holds if and only if $n=0$. This implies that for the same array gain factor $M$, the proposed RAA can provide finer angular resolution than the conventional ULA. Moreover, the beamwidth of the RAA  remains constant across all signal directions, while that for the conventional ULA becomes wider as the signal moves away from the boresight direction. Therefore, compared to the ULA with the same array gain factor $M$, the RAA can offer finer and uniform angular resolution for the total angle coverage range.
    \item \emph{Enhanced beamforming gain}: By comparing (\ref{2DRAA_AP}) and (\ref{ULA_AP}),
        it is found that to achieve the same coverage range $[-\phi_{\max}, \phi_{\max}]$, the beamwidth of antenna elements in the proposed RAA can be much narrower than that in the conventional ULA, i.e.,
        \begin{equation}
        \frac{\phi_{\text{3dB}}^{\text{ULA}}}{\phi_{\text{3dB}}}=\frac{2\phi_{\text{max}}}{\arcsin(2/M)}\approx\frac{2\phi_{\text{max}}}{2\phi_{\text{max}}/(N-1)}=N-1>1.
         \nonumber
        \end{equation}
         Thus, based on (\ref{G00}), a further comparison of the peak gains between the RAA and ULA with  $G_{\text{sum}}=G^{\text{ULA}}_{\text{sum}}$ gives the following result
        \begin{equation}
          \frac{G(0)}{G_{\text{ULA}}(0)} =\frac{\phi_{\text{3dB}}^{\text{ULA}}}{\phi_{\text{3dB}}}>1.
           \nonumber
        \end{equation}
        The narrower beamwidth of antenna elements in the RAA implies stronger directivity.
      Therefore, for the same $M$, RAA can be designed to provide enhanced overall beamforming gain compared to the conventional ULA.
    \item \emph{Dramatically reduced hardware costs}: In contrast to the conventional ULA with HBF that requires $MN_{\text{RF}}$ phase shifters and $M$ antenna elements, the proposed RAA requires $N_{\text{RF}}N$ RF switches and $NM$ antenna elements. Accordingly, the primary hardware expenditures for the RAA and ULA architectures can be expressed as
$\text{cost}_{\text{RAA}}=N_{\text{RF}}Np_{\text{sw}}+NMp_{\text{ant}}$ and $
\text{cost}_{\text{ULA}}=N_{\text{RF}}Mp_{\text{ps}}+Mp_{\text{ant}}$,
 respectively, where $p_{\text{sw}}$, $p_{\text{ant}}$, and $p_{\text{ps}}$ denote the unit prices of RF switch, antenna element, and phase shifter, respectively.
  For commercial components in high-frequency systems, $p_{\text{ps}}$, $p_{\text{sw}}$, and $p_{\text{ant}}$ are typically on the order of $100 \$$, $10 \$$, and $0.01 \$$, respectively \cite{qorvo2025}. Therefore, RAA is expected to  significantly reduce hardware costs, i.e., $
  \text{cost}_{\text{RAA}}< \text{cost}_{\text{ULA}}$.
For example, we consider $M=128$, $N_{\text{NF}}=16$, and $\phi_{\max}=0.499\pi$, which yields $N=201$.
The wireless system operates at the carrier frequency of $38$ GHz and employs the TGP2102 5-bit digital phase shifter and the TGS4302 RF switch, both manufactured by Qorvo \cite{qorvo2025}.
According to the manufacturer quotations, the unit prices are roughly $p_{\text{ps}}\approx131.2 \$$, $p_{\text{sw}}\approx14.31 \$$ (for single-pole single-throw), and $p_{\text{ant}} \approx 0.01 \$$.
 Consequently, the hardware costs are $\text{cost}_{\text{RAA}} =46278 \$$ and $\text{cost}_{\text{ULA}}=268700 \$$.
Therefore, the proposed RAA can significantly reduce hardware cost, amounting to only $17.22\% $ of that required by the ULA-HBF architecture. 
\end{itemize}

\begin{figure}[htbp]
\centering
\includegraphics[width=0.7\linewidth]{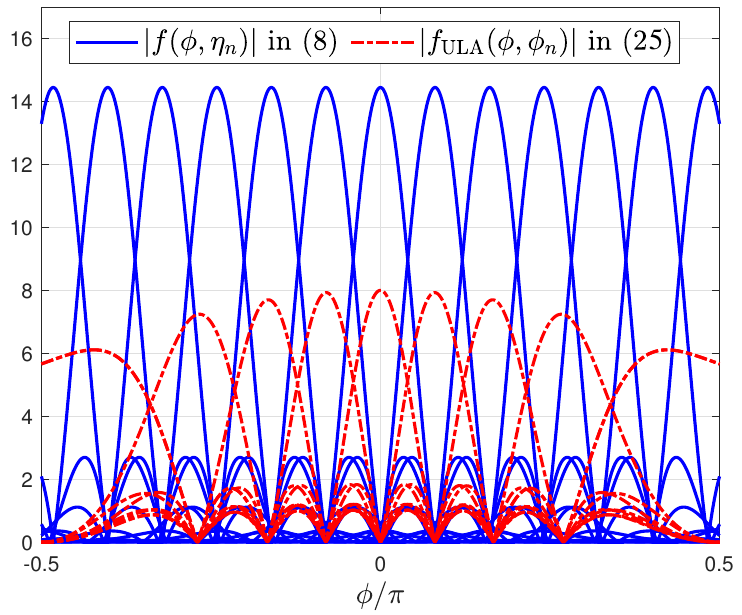}
	\caption{Beam patterns $|f(\phi,\eta_n)|$ in (\ref{steering_eff00}) and $|f_{\text{ULA}}(\phi,\phi_n)|$ in (\ref{steeringV1X}), where $M=8$ and $\phi_{\text{max}}=0.499\pi$.}
	\label{2DRAA_2}
\end{figure}

The beam patterns of the proposed RAA $|f(\phi,\eta_n)|$ in (\ref{steering_eff00}) and the conventional ULA $|f_{\text{ULA}}(\phi,\phi_n)|$ in (\ref{steeringV1X}) are illustrated in Fig. \ref{2DRAA_2},  where $M=8$ and $\phi_{\text{max}}=0.499\pi$.
The antenna element pattern follows the 3GPP model in (\ref{G}), where we set $\phi_{\text{3dB}}=0.3\pi$ and $G(0)=5.13$ dB for the RAA, and $\phi_{\text{3dB}}^{\text{ULA}}=\pi$ and $G_{\text{ULA}}(0)=0$ dB for the ULA, ensuring that $G_{\text{sum}}=G^{\text{ULA}}_{\text{sum}}$.
Fig. \ref{2DRAA_2} verifies that the proposed RAA achieves finer and uniform angular resolution across the total angle coverage range, and enhanced beamforming gain, compared to the conventional ULA.


\section{RAA-based Uplink Multi-user Communications}

In this section, we apply the proposed RAA structure for uplink multi-user communications.
In particular, the RAA beamforming and ray selection are performed based on the channel vector $\vec{h}_k$ in (\ref{channel modelX}). In a multi-path environment, the effective channel $\vec{h}_k$ for the RAA-based communications is given by
\begin{equation}\label{h_up}
\begin{aligned}
\vec{h}_k=\sum\nolimits_{l =1}^{L_k}\alpha_{k,l} \vec{f}(\phi_{k,l}),
\end{aligned}
\end{equation}
where $L_k$ represents the number of propagation paths for user $k$;
$\alpha_{k,l}$ denotes the complex gain of the $l$th path for user $k$;  
$\phi_{k,l}\in \left[-\phi_{\text{max}},\phi_{\text{max}}\right]$ is the $l$th path angle for user $k$;
and $\vec{f}(\phi_{k,l})\in \mathbb{C}^{N\times 1}$ denotes the resulting outputs of the $N$ sULAs for path angle $\phi_{k,l}$.
Noted that the channel vector for the $M$-element ULA, denoted by $\vec{h}^{\text{ULA}}_k\in \mathbb{C}^{M\times 1}$, is expressed as
 \begin{equation}
\begin{aligned}
\vec{h}^{\text{ULA}}_k&=\sum\nolimits_{l =1}^{L_k}\alpha_{k,l} \vec{a}_{\text{ULA}}(\phi_{k,l}).
\end{aligned}
\end{equation}

It is noted that the noise at the $M$-element ULA, denoted by $\vec{z}'\in \mathbb{C}^{M\times 1}$, follows i.i.d. CSCG distribution with variance $\sigma^2$, i.e., $\vec{z}'\sim N_{\mathbb{C}}\left(0,\sigma^2\vec{I}_{M}\right)$.
Accordingly, the noise at each sULA $n$, denoted by $\vec{z}'_n\in \mathbb{C}^{M\times 1}$, is modeled as $\vec{z}'_n\sim N_{\mathbb{C}}\left(0,\sigma^2\vec{I}_{M}\right)$, and is assumed to be independent across different sULAs.
Thus, the effective noise at the $N$ sULAs can be expressed as
 $\vec{z}=[\vec{1}_{M\times 1}^T\vec{z}'_n]_{n\in\mathcal{N}}\sim N_{\mathbb{C}}\left(0,\sigma_0^2\vec{I}_{N}\right)$.  
 Therefore, the relationship between noise variance at the RAA and ULA is $\sigma_0^2=M\sigma^2$.

In the special case of single-user communication, for any given ray selection matrix $\vec{S}$, the optimal baseband beamforming in (\ref{SNR}) is the
MRC, i.e., $\vec{w}=\vec{S}\vec{h}$.
Accordingly, the resulting SNR in (\ref{SNR}) can be rewritten as a function of
$\vec{S}$ as follows
 \begin{equation}\label{SNR2}
\text{SNR}(\vec{S})=\bar{P_t}\|\vec{S}\vec{h}\|^2/M,
 \end{equation}
where $\bar{P_t}=\frac{P_t}{\sigma^2}$ represents the transmit SNR.
 It is observed from (\ref{SNR2}) that, to maximize the SNR, the optimal $\vec{S}$ corresponds to selecting the $N_{\text{RF}}$ entries of $\mathbf h$ with the largest magnitudes.
Specifically, in the case of a single path, the $n$th entry of $\vec{h}$ simplifies to $h_n=M\alpha\sqrt{G(0)}$ when  $\phi=\eta_n$. Thus, the resulting maximum SNR in (\ref{SNR2}) is $M\bar{P_t}|\alpha|^2G(0)$, which increases linearly with the array gain factor $M$. 

For the more general case of multi-user communications,
the problem of maximizing sum rate is formulated as a joint optimization of the receive beamforming matrix $\vec{W}=\left[\vec{w}_k\right]_{1\leq k \leq K}\in \mathbb{C}^{ N_{\text{RF}}\times K}$ and ray selection matrix $\vec{S}$, as follows
\begin{equation}\label{channel model4}
\begin{aligned}
   \underset{\vec{W},\vec{S}}{\text{max}}& \quad R_{\text{sum}}=\sum\nolimits_{k=1}^{K}\text{log}_2(1+\text{SINR}_k)\\
   \text{s.t.}& \quad(\text{C1}):
               \enspace\vec{S}\in \{0,1\}^{N_{\text{RF}}\times N},\\
               &\quad(\text{C2}):\enspace\left\|[\vec{S}]_{i,:}\right\|_0=1,\enspace 1\leq i\leq N_{\text{RF}},\\
               &\quad(\text{C3}):\enspace\left\|[\vec{S}]_{:,n}\right\|_0\leq 1,\enspace 1\leq n\leq N.\\
   \end{aligned}
\end{equation}

Solving Problem (\ref{channel model4}) directly is challenging due to the binary constraints in (C1) and the non-convex $l_0$-norm in (C2)-(C3).
Moreover, the beamforming matrix $\vec{W}$ is tightly coupled with the ray selection matrix $\vec{S}$, further complicating the optimization problem.
Although an exhaustive search over all possible combinations of selecting $N_{\text{RF}}$ rays from $N$ sULAs could yield the optimal solution, the search space size $\tbinom{N}{N_{\text{RF}}}$ becomes computationally infeasible for large $N$ and ${N_{\text{RF}}}$.
To address this issue, we propose a low-complexity greedy algorithm that reduces the computational complexity to $O(NN_{\text{RF}})$.

For a given ray selection matrix $\vec{S}$, the beamforming matrix that maximizes the sum rate in (\ref{channel model4}) is known to be the classic minimum mean square error (MMSE) beamforming:
\begin{equation}\label{203}
\vec{w}_k(\vec{S})=\vec{C}_k^{-1}(\vec{S})\vec{S}\vec{h}_k,
\end{equation}
 with $\vec{C}_k(\vec{S})\triangleq\vec{S} \bigl(\sum_{i \neq k}\vec{h}_i\vec{h}_i^H+\frac{M}{\bar{P_t}}\vec{I}_N\bigl)\vec{S}^H$ being the interference-plus-noise covariance matrix.
Then, by substituting the matrices $\vec{S}$ and $\vec{W}(\vec{S})$ into (\ref{channel model4}), the resulting maximum sum rate associated with the selected ray index set, denoted by $\Omega=\{n:\left\|[\vec{S}]_{:,n}\right\|_0=1\}$, can be expressed as
\begin{equation}\label{channel model5}
    R_{\text{sum}}(\Omega)=\sum\nolimits_{k=1 }^{K}\text{log}_2\bigl(1+(\vec{S}\vec{h}_k)^H \vec{C}_k^{-1}(\vec{S})\vec{S}\vec{h}_k\bigl).
\end{equation}

In the proposed greedy algorithm, one sULA is selected at each iteration to maximize the sum rate.
 Specifically, at the $i$th iteration, the optimal ray index $n^{(i)}$ is selected as
$n^{(i)}=\underset{n}{\text{argmax}} \quad R_{\text{sum}}(\Omega^{(i-1)}\cup n)$,
and the ray index set is updated as $\Omega^{(i)}=\Omega^{(i-1)}\cup n^{(i)}$, with the corresponding sum rate calculated by $R_{\text{sum}}(\Omega^{(i)})$ in (\ref{channel model5}).
This iterative procedure continues until a total of $N_{\text{RF}}$ sULAs are selected.
The proposed greedy scheme for uplink joint RAA beamforming and ray selection is summarized in Algorithm 1.

\begin{algorithm}[htb]
\caption{The proposed greedy algorithm for uplink RAA beamforming and ray selection}
\label{alg:Framwork}
\begin{algorithmic}[1] 
\STATE Initialize $\Omega^{(0)}=\emptyset$ and $\mathcal{N}=\{1,2,\cdots,N\}$.\\ 
\STATE \textbf{for} $i=1:N_{\text{RF}}$ \textbf{do}\\ 
\STATE  \quad Calculate $n^{(i)}=\underset{n\in \mathcal{N}}{\text{argmax}} \quad R_{\text{sum}}(\Omega^{(i-1)}\cup n)$;
\STATE \quad Update $\Omega^{(i)}=\Omega^{(i-1)}\cup n$ and $\mathcal{N}=\mathcal{N}\backslash n^{(i)}$;
\STATE \textbf{end for}
\STATE Obtain $\vec{S}$ based on the index set $\Omega^{(N_{\text{RF}})}$;
\STATE Calculate $\vec{W}$ with (\ref{203}) and $R_{\text{sum}}$ with (\ref{channel model5});
\STATE \textbf{Output:} $\vec{W}$, $\vec{S}$, and $R_{\text{sum}}$.
\end{algorithmic}
\end{algorithm}
\vspace{-0.2cm}

\section{RAA-based Downlink Multi-user Communications}
In RAA-based downlink multi-user communication system,
the RF combiner of each sULA needs to be replaced by power splitter.
The downlink transmit information-bearing symbol vector is denoted by
$\vec{s}_{\text{DL}}=[s_{\text{DL},k}]_{1 \leq k\leq K}\in\mathbb{C}^{K\times 1}$, satisfying $\mathbb{E}[\vec{s}_{\text{DL}}\vec{s}^H_{\text{DL}}]=\vec{I}_{K}$.
The downlink baseband transmit beamforming matrix is denoted by $\vec{W}_{\text{DL}}=\left[\vec{w}_{\text{DL},k}\right]_{1\leq k \leq K}\in \mathbb{C}^{ N_{\text{RF}}\times K}$ with $\vec{w}_{\text{DL},k}\in\mathbb{C}^{ N_{\text{RF}}\times 1}$ being the transmit beamforming vector for user $k$. Thus, after undergoing the baseband processing and RSN, the transmit signal vector at the $N$ sULAs, denoted by $\vec{x}_{\text{DL}}\in\mathbb{C}^{N\times1}$, is given by
\begin{equation}
\vec{x}_{\text{DL}}=\vec{S}^H\vec{W}_{\text{DL}}\vec{s}_{\text{DL}}=\vec{S}^H\sum\nolimits_{k=1}\nolimits^{K}\vec{w}_{\text{DL},k}s_{\text{DL},k},
\end{equation}
where $\vec{S}\in\{0,1\}^{N_{\text{RF}}\times N}$ is the RSN matrix, and the transmit signal vector satisfies $\mathbb{E}[\|\vec{x}_{\text{DL}}\|^2]\leq P_{\text{DL}}$, with $P_{\text{DL}}$ being the downlink transmit total power. Thus, we have $\mathbb{E}[\|\vec{x}_{\text{DL}}\|^2]=\mathbb{E}[\|\vec{W}_{\text{DL}}\vec{s}_{\text{DL}}\|^2]=\sum_{k=1}^{K}\|\vec{w}_{\text{DL},k}\|^2\leq P_{\text{DL}}$, where the first equality holds due to $\vec{S}\vec{S}^H=\vec{I}_{N_\text{RF}}$.
Accordingly, the downlink received signal at user $k$ can be formulated as
\begin{equation}\label{hX}
    y_{\text{DL},k}=\left(\vec{h}^{\text{act}}_{{\text{DL}},k}\right)^H \left(\vec{x}_{\text{DL}}\otimes \bigl(\sqrt{1/M}\vec{1}_{M\times1}\bigl)\right)
    +z_k,
\end{equation}
where $\vec{h}^{\text{act}}_{{\text{DL}},k}\in\mathbb{C}^{NM\times1}$ denotes the channel vector between all $NM$ antenna elements of RAA and user $k$, modeled as
\begin{equation}\label{h_DL}
\begin{aligned}
\vec{h}^{\text{act}}_{{\text{DL}},k}=\sum\nolimits_{l =1}^{L_k}\alpha_{k,l}\text{vec}\left(\vec{A}(\phi_{k,l})\right),
\end{aligned}
\end{equation}
 and $z_k\sim N_{\mathbb{C}}(0,\sigma^2)$ is the noise at user $k$, following CSCG distribution with variance $\sigma^2$.
By substituting (\ref{h_DL}) into (\ref{hX}), we have
\begin{equation}\label{hX0}
\begin{aligned}
    y_{\text{DL},k}
    &=\sqrt{1/M}\underbrace{\Bigl(\sum\nolimits_{l =1}^{L_k}\alpha_{k,l}\vec{f}(\phi_{k,l})\Bigl)^H}_{\vec{h}_{{\text{DL}},k}^H}\vec{x}_{\text{DL}}+z_k\\
    &=\sqrt{1/M}\vec{h}_{{\text{DL}},k}^H\vec{S}^H\sum\nolimits_{i=1}\nolimits^{K}\vec{w}_{\text{DL},i}s_{\text{DL},i}+z_k,
\end{aligned}
\end{equation}
where $\vec{h}_{{\text{DL}},k}\in\mathbb{C}^{N\times 1}$ is the effective channel between the $N$ sULA ports and user $k$, which is  identical to (\ref{h_up}). 
Based on (\ref{hX0}), the SINR of user $k$ for the downlink RAA-based communications is given by
\begin{equation}\label{channel modelXX3}
     \text{SINR}_{\text{DL},k}=\frac{\frac{1}{M}\bigl|\vec{h}_{\text{DL},k}^H\vec{S}^H\vec{w}_{\text{DL},k}
    \bigl|^2}{\sum_{i\neq k}\frac{1}{M}\bigl|\vec{h}_{\text{DL},k}^H\vec{S}^H\vec{w}_{\text{DL},i}
    \bigl|^2+\sigma^2}.
\end{equation}

In particular, the SNR for downlink single-user communications is
 \begin{equation}\label{DLSNR}
\text{SNR}_{\text{DL}}=\left|\vec{w}_{\text{DL}}^H\vec{S}\vec{h_{\text{DL}}}
    \right|^2/(M\sigma^2).
 \end{equation}

By observing (\ref{DLSNR}), the design of downlink joint RAA beamforming and ray selection is similar to the uplink counterpart.
Specifically, for a given ray selection matrix $\vec{S}$, the optimal baseband beamforming in (\ref{DLSNR}) is the MRT, i.e., $\vec{w}_{\text{DL}}(\vec{S})=\sqrt{P_{\text{DL}}}\frac{\vec{S}\vec{h_{\text{DL}}}}{\|\vec{S}\vec{h}_{\text{DL}}\|}$.
Accordingly, (\ref{DLSNR}) can be rewritten as a function of $\vec{S}$
 \begin{equation}\label{DL_SNR2}
\text{SNR}_{\text{DL}}(\vec{S})=\bar{P}_{\text{DL}}\|\vec{S}\vec{h}_{\text{DL}}\|^2/M,
 \end{equation}
where $\bar{P}_{\text{DL}}=\frac{P_{\text{DL}}}{\sigma^2}$ denotes the downlink transmit SNR.
Similar to the uplink case, the optimal $\vec{S}$ that maximizes the SNR also corresponds to selecting the $N_{\text{RF}}$ entries of $\mathbf h_{\text{DL}}$ with the largest magnitudes. Moreover, the downlink maximum SNR in (\ref{DL_SNR2}) is similar to the uplink counterpart in (\ref{SNR2}).

For downlink multi-user RAA-based communications,
the problem of maximizing the minimum SINR is formulated as a joint optimization of $\vec{W}_{\text{DL}}$ and $\vec{S}$, as follows
\begin{equation}\label{SINR2}
\begin{aligned}
   \underset{\vec{W_{\text{DL}}},\vec{S}}{\text{max}}& \enspace \bigl(\underset{\forall k}{\text{min}}  \enspace\text{SINR}_{\text{DL},k}\bigl)\\
   \text{s.t.}& \quad(\text{C1}):
               \enspace\vec{S}\in \{0,1\}^{N_{\text{RF}}\times N},\\
               &\quad(\text{C2}):\enspace\left\|[\vec{S}]_{i,:}\right\|_0=1,\enspace 1\leq i\leq N_{\text{RF}},\\
               &\quad(\text{C3}):\enspace\left\|[\vec{S}]_{:,n}\right\|_0\leq 1,\enspace 1\leq n\leq N,\\
               &\quad(\text{C4}):\enspace\sum\nolimits_{k=1}\nolimits^{K}\|\vec{w}_{\text{DL},k}\|^2\leq P_{\text{DL}}.
   \end{aligned}
\end{equation}

Problem (\ref{SINR2}) is difficult to solve directly due to binary constraints and the tight coupling between $\vec{W}_{\text{DL}}$ and $\vec{S}$.
To address this issue, we propose an alternating optimization algorithm for downlink joint RAA beamforming and ray selection. 
We first fix $\vec{S}$, and then reformulate Problem (\ref{SINR2}) by introducing an auxiliary variable $\gamma$ to replace the minimum SINR objective, as follows
\begin{equation}\label{SINR4}
\begin{aligned}
   \underset{\vec{W}_{\text{DL}},\gamma}{\text{max}}& \enspace \gamma\\
   \text{s.t.}
               &\quad(\text{C4}) \text{ in Problem } (\ref{SINR2}),\\
    &\quad\text{SINR}_{\text{DL},k}\geq \gamma, \forall k.\\
   \end{aligned}
\end{equation}

Problem in (\ref{SINR4}) can be efficiently solved by using bisection search and second-order cone programming (SOCP) to find the maximum feasible $\gamma^*$ and corresponding feasible $\vec{W}_{\text{DL}}^*$.
 During the bisection search process, for the current $\gamma$, we solve the SOCP problem to determine whether a feasible solution $\vec{W}_{\text{DL}}$ exists. Then, based on the obtained feasible $\vec{W}_{\text{DL}}^*$, Problem (\ref{SINR2}) is reformulated as
 \begin{equation}\label{SINR5}
\begin{aligned}
   \underset{\vec{S}}{\text{max}} &\enspace \bigl(\underset{\forall k}{\text{min}}  \enspace\text{SINR}_{\text{DL},k}\bigl)\\
   \text{s.t.}& \quad(\text{C1})-(\text{C3}) \text{ in Problem } (\ref{SINR2}).
   \end{aligned}
\end{equation}

Exhaustive search is utilized to obtain the optimal solution $\vec{S}^*$ in Problem (\ref{SINR5}).
The above process continues until $\gamma^*$ converges. The proposed alternating optimization algorithm is summarized in Algorithm 2.

\begin{algorithm}[htb]
\caption{Alternating optimization algorithm for downlink RAA beamforming and ray selection}
\label{alg:Alternating}
\begin{algorithmic}[1]
\STATE \textbf{Input:} maximum iteration $T_{\text{max}}$, convergence threshold $\varepsilon_{\text{DL}}$.
\STATE Initialize $\mathbf{S}^{(0)}$, $\gamma^{(0)}=0$, $t=0$.
\REPEAT
    \STATE  Fix $\mathbf{S}^{(t)}$, use bisection and SOCP to solve Problem (\ref{SINR4}) to obtain maximum feasible $\gamma^{(t+1)}$ and $\mathbf{W}_{\text{DL}}^{(t+1)}$;
    \STATE  Fix $\mathbf{W}_{\text{DL}}^{(t+1)}$, use exhaustive search to solve Problem (\ref{SINR5}) to obtain the optimal $\mathbf{S}^{(t+1)}$;
    \IF {$|\gamma^{(t+1)} - \gamma^{(t)}| \leq \varepsilon_{\text{DL}}$ or $t+1 \geq T_{\text{max}}$}
        \STATE \quad break
    \ENDIF
    \STATE $t \leftarrow t + 1$
\UNTIL{convergence}
\STATE \textbf{Output:} $\mathbf{W}_{\text{DL}}^{(t+1)}$, $\mathbf{S}^{(t+1)}$, and $\gamma^{(t+1)}$.
\end{algorithmic}
\end{algorithm}

\section{Simulation Results}
In this section, simulation results are presented to evaluate communication performance of the proposed RAA architecture.
Unless otherwise specified, we set $M=128$ and $\phi_{\text{max}}=0.499\pi$,
resulting in $N=201$ and $N'=129$.
In addition, we set $N_{\text{RF}}=8$ and $K=8$. 
Both the uplink transmit SNR $\bar{P}_t$ and downlink transmit SNR $\bar{P}_{\text{DL}}$ are varied from -10 dB to 10 dB. 
For the antenna element pattern, we set $\phi_{\text{3dB}}=0.3\pi$ and $G(0)=5.1335$ dB for the proposed RAA, and $\phi_{\text{3dB}}^{\text{ULA}}=\pi$ and $G_{\text{ULA}}=0$ dB for the conventional ULA,
ensuring that $G_{\text{sum}}=G_{\text{sum}}^{\text{ULA}}$.
In addition, we also consider isotropic antenna pattern whose
gain is set to –2.816 dB in all directions to ensure the same total power gain as the directional antenna element.
The UMa NLoS scenario at the carrier frequency of 47.2 GHz is considered, where the channel parameters are defined in Table \ref{tab:cluster_power_model}. The number of channel realizations is set to 50.
For the proposed Algorithm 2, we set $T_{\text{max}}=20$ and $\varepsilon_{\text{DL}}=10^{-3}$.

\begin{table*}[htbp]
\centering
\caption{Channel Parameters for UMa NLoS Scenario \cite{3GPPchannel}. }
\begin{tabular}{|c|c|c|}
\hline
\textbf{Parameter} & \textbf{Symbol} & \textbf{Definition / Value} \\
\hline
Carrier frequency in GHz & $f_c$ & $47.2$ GHz \\\hline
Number of clusters & $N_c$ & 12 \\\hline
Rays per cluster & $N_r$ & $20$ \\\hline
angle spread &AS &  \makecell{$\text{lgAS}\sim N(2.08-0.27\text{log}_{10}(f_c),0.11^2)$, lgAS=$\text{log}_{10}(\mathrm{AS}/1^{\circ})$
}\\\hline


Power of cluster $n$  &$P_n$ &$ P_n=\frac{P'_n}{\sum_{n=1}^{N_c}P'_n}\text{ with }
P'_n =  10^{-\frac{Z_n}{10}}, Z_n\sim N(0,3^2)$\\\hline

 Power of ray $m$ in cluster $n$  &$P_{n,m}$ &\makecell{ $P_{n,m}=P_n\frac{P'_{n,m}}{\sum_{m=1}^{N_r}P'_{n,m}}\text{ with } P'_{n,m} = e^{-\frac{\sqrt{2}|x_{n,m}|}{15} }$, $x_{n,m}\sim \text{Unif}(-2,2)$
}\\\hline

Path gain & $\alpha_{n,m}$ & $\alpha_{n,m} = \sqrt{P_{n,m}} \cdot e^{j \varphi_{n,m}}, \quad \varphi_{n,m} \sim \mathrm{Unif}(-\pi, \pi)$ \\\hline
 Angle of cluster $n$ & $\phi_n$ &  \makecell{$\phi_{n} = X_n \frac{2(\text{AS}/1.4)\sqrt{-\ln(P_n/\max(P_n))}}{1.289} + Y_n + \phi_{\text{LOS}}$\\$X_n$ is uniformly distributed to the set $\{1, -1\}$, $Y_n\sim N(0,(\text{AS}/7)^2)$, $\phi_{\text{LOS}}\sim \text{Unif}(-90^\circ,90^\circ)$}\\\hline

 Angle of ray $m$ in cluster $n$ & $\phi_{n,m}$ & $\phi_{n,m}=\phi_{n}+15\alpha_m$,  $\alpha_m\sim\text{Unif}(-2,2)$ \\\hline


\end{tabular}
\label{tab:cluster_power_model}
\end{table*}

\begin{figure}[htbp]
\centering
	\includegraphics[width=0.7\linewidth]{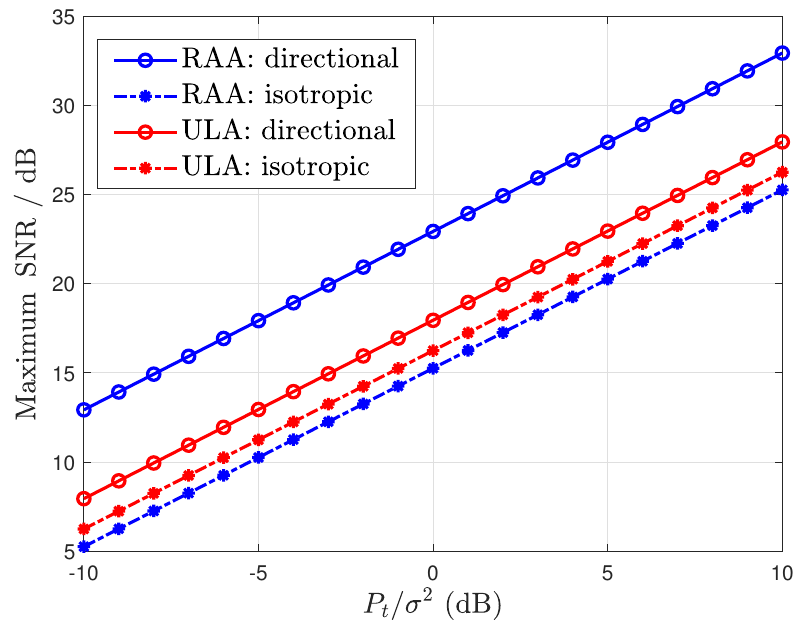}
\caption{Single-user uplink maximum $\text{SNR}$ (dB) in (\ref{SNR2}) versus $\bar{P}_t$ for the proposed RAA and the conventional ULA.} 
\label{Single_1}
\end{figure}

Fig. \ref{Single_1} shows the single-user uplink maximum $\text{SNR}$ (dB) in (\ref{SNR2}) versus transmit SNR $\bar{P}_t$ for the proposed RAA and the conventional ULA with HBF. 
It can be observed from Fig. \ref{Single_1} that, when the directional antenna elements are employed, RAA achieves an SNR approximately 5 dB higher than that of the ULA across different transmit SNR levels.
This performance improvement is attributed to the enhanced beamforming gain of the RAA.
In contrast, when isotropic antennas are used, ULA slightly outperforms RAA, as RAA's finer and uniform angular resolution makes it more challenging to align with the signal directions.
These results demonstrate that the proposed RAA can enhance the uplink single-user communication performance compared to the ULA.

\begin{figure}[htbp]
\centering
\subfigure[]{
\includegraphics[width=0.68\linewidth]{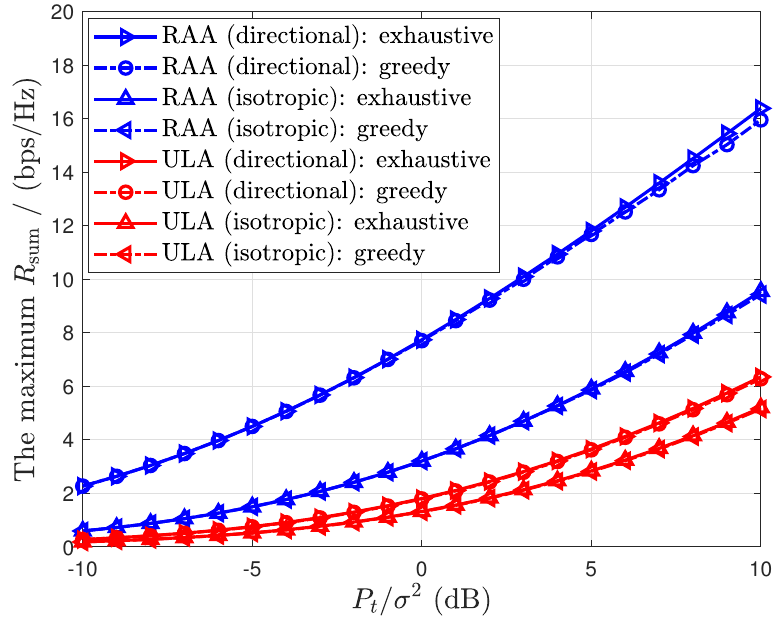}}
\subfigure[]{
	\includegraphics[width=0.68\linewidth]{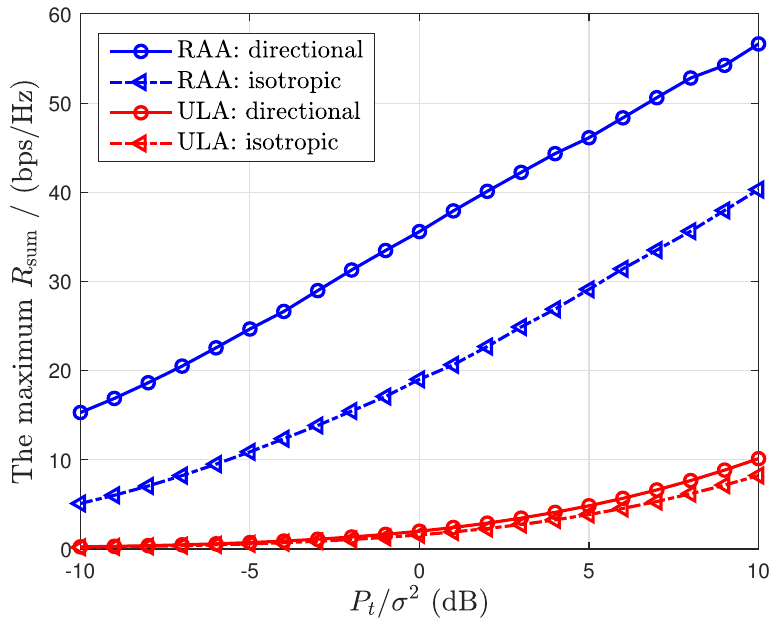}}
\caption{Multi-user uplink maximum sum rate $R_{\text{sum}}$ (bps/Hz) in (\ref{channel model4}) versus $\bar{P}_t$ for the RAA and ULA, where (a) $M=6$, $N_{\text{RF}}=3$ and $K=3$, (b) $M=128$, $N_{\text{RF}}=8$ and $K=8$.} 
\label{multiple_1}
\end{figure}

Fig. \ref{multiple_1} illustrates the multi-user uplink communication sum rates $R_{\text{sum}}$ (bps/Hz) in (\ref{channel model4}) versus transmit SNR $\bar{P}_t$ for the proposed RAA and the conventional ULA. 
Due to the prohibitive complexity of exhaustive search, a special case with $M=6$, $N_{\text{RF}}=3$, and $K=3$ is considered in Fig. \ref{multiple_1} (a), where $R_{\text{sum}}$ is obtained using both exhaustive search and the proposed Algorithm 1. In Fig. \ref{multiple_1} (b), only the greedy scheme is considered, where $M=128$, $N_{\text{RF}}=8$, and $K=8$.
As shown in Fig. \ref{multiple_1} (a), $R_{\text{sum}}$ achieved by the proposed Algorithm 1 achieves the near-optimal performance as exhaustive search, validating the effectiveness of the proposed greedy scheme.
Furthermore, Fig. \ref{multiple_1} demonstrates that the RAA significantly outperforms the ULA for multi-user  communication, especially with directional antennas.
This improvement is attributed to the higher beamforming gain as well as the finer and uniform angular resolution of the proposed RAA.

\begin{figure}[htbp]
\centering
	\includegraphics[width=0.68\linewidth]{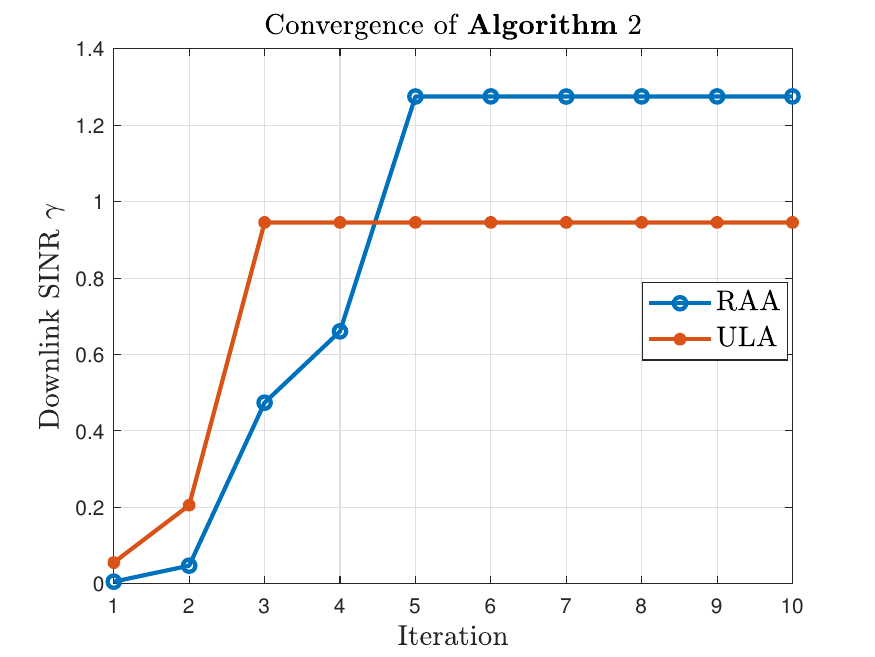}
\caption{Illustration of the convergence behavior of the proposed Algorithm 2, where $M = 6$, $N_{\text{RF}} = 3$, $K = 3$, and  $\bar{P}_{\text{DL}}=10$ dB.} 
\label{Coverageance}
\end{figure}

Fig. \ref{Coverageance} illustrates the convergence behavior of the proposed Algorithm 2 for the RAA and ULA, where $M = 6$, $N_{\text{RF}} = 3$, $K = 3$, and $\bar{P}_{\text{DL}}=10$ dB.
For the ULA case, Algorithm 2 is utilized to select the best $N_{\text{RF}}$ DFT codewords and design the optimal baseband beamforming for downlink max-min SINR.
As shown in Fig. \ref{Coverageance},
the proposed Algorithm 2 converges rapidly within a few iterations for both RAA and ULA architectures. 
 These results validate the convergence and effectiveness of the proposed alternating optimization algorithm.

\begin{figure}[htbp]
\centering
	\includegraphics[width=0.68\linewidth]{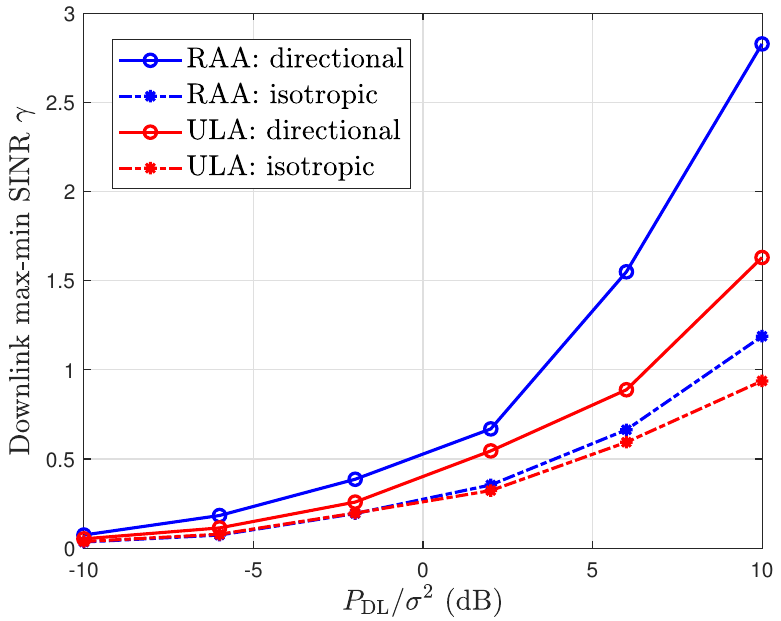}
\caption{Multi-user downlink max-min SINR $\gamma$ in (\ref{SINR2}) versus transmit SNR $\bar{P}_{\text{DL}}$ for the proposed RAA and the conventional ULA with HBF, where $M = 6$, $N_{\text{RF}} = 3$, and $K = 3$.} 
\label{DL_multiuser}
\end{figure}

Fig. \ref{DL_multiuser} depicts the multi-user downlink max-min SINR $\gamma$ in (\ref{SINR2}) versus the transmit SNR $\bar{P}_{\text{DL}}$ for the proposed RAA and the conventional ULA with HBF. 
It is evident that the RAA with directional antenna elements significantly outperforms all other configurations, particularly in the high-SNR regime, owing to its enhanced beamforming gain.
In the case of isotropic antenna elements,
RAA can still improve communication performance at high transmit SNR, thanks to its finer and uniform angular resolution over the entire angle coverage range.
These results further confirm the advantage of the proposed RAA in improving downlink communication quality and interference suppression compared to conventional ULA architecture.

\section{Conclusion}
In this paper, we proposed a novel multi-antenna architecture, termed RAA, to enable flexible and cost-effective beamforming for high-frequency wireless systems.
Compared to the conventional ULA with HBF,  the proposed RAA can achieve finer and uniform angular resolution, enhanced beamforming gain, and substantial hardware cost reduction, though requires a larger size.
To optimize uplink and downlink RAA-based communications for both single-user and multi-user scenarios, efficient joint ray selection and beamforming algorithms were proposed. Simulation results validated that the proposed RAA outperforms conventional ULA with HBF in terms of communication performance, while significantly reducing implementation costs.
These findings highlight the potential of the RAA as a practical and scalable solution for future high-frequency wireless communication and sensing applications.
In future work, we will address the issue of antenna blockage in practical RAA systems and further explore the design of three-dimensional (3D) RAA architecture.

\begin{appendices}
\section{ PROOF OF THEOREM \ref{lemmaX0001}}\label{AA}
\begin{figure}[htbp]
\centering
\includegraphics[width=0.66\linewidth]{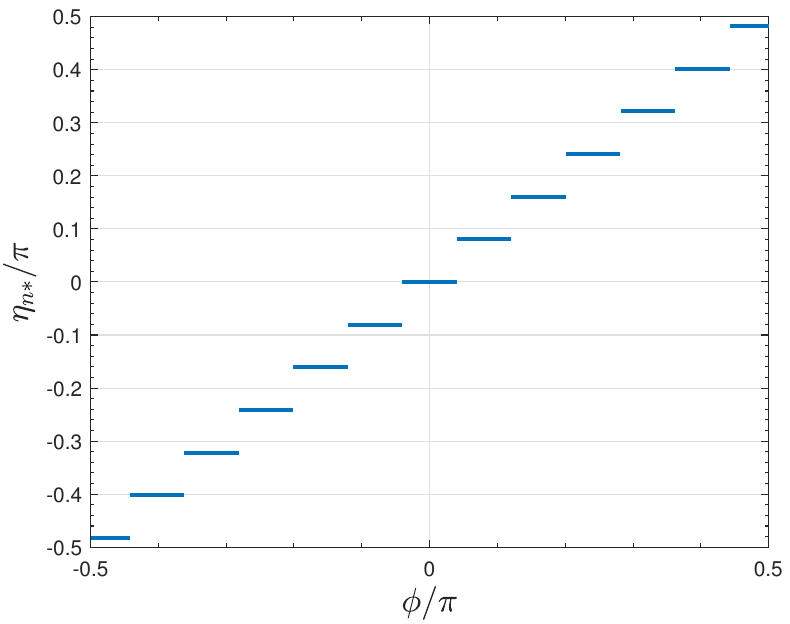}
	\caption{The angle $\eta_{n^*}=n^*\times\arcsin(2/M)$ with $n^*$ being the optimal solution to (\ref{we}) vs any given signal direction $\phi$.}
	\label{fenduan}
\end{figure}

Under the two assumptions (i) and (ii) stated in Theorem 1, substituting (\ref{steering_bn}) and (\ref{steering_eff00}) into (\ref{we}) leads to a reformulation that holds under the condition $\phi_{\text{3dB}}\geq\arcsin(2/M)$
\begin{equation}\label{we1}
\begin{aligned}
f(\phi)=M\sqrt{G(\phi-\eta_{n^*})} \left|H_M(\sin(\phi-\eta_{n^*}))\right|,
\end{aligned}
\end{equation}
where $\eta_{n^*}=n^*\times\arcsin\left(\frac{2}{M}\right)$ with $n^*=\bigl\lfloor\bigl(\bigl\lfloor\frac{2\phi}{\arcsin(2/M)}\bigl\rfloor+1\bigl)/2\bigl\rfloor$ being the optimal solution to (\ref{we}).
The angle $\eta_{n^*}$ vs any given signal direction $\phi$ is illustrated in Fig. \ref{fenduan}, $\forall n^*\in\mathcal{N}$.
It is observed from Fig. \ref{fenduan} that when $\phi\in \Omega_{n^*}$ with $\Omega_{n^*}=\{\zeta:\text{max}\left((n^*-0.5)\arcsin\left(\frac{2}{M}\right),-\phi_{\text{max}}\right)\leq\zeta <\text{min}\left((n^*+0.5)\arcsin\left(\frac{2}{M}\right),\phi_{\text{max}}\right)\}$, the optimal solution $n^*$ remains unchanged, $\forall n^*\in\mathcal{N}$.
Since the function $f(\phi)$ in (\ref{we1}) increases linearly w.r.t. the array gain factor $M$, we re-express the threshold as $\varepsilon_0=M\varepsilon$. Therefore, by substituting (\ref{we1}) and $\varepsilon_0=M\varepsilon$ into (\ref{we01}), the antenna element pattern of RAA $G(\zeta)$ needs to designed to satisfy
\begin{equation}\label{we02}
\underset{-\phi_{\text{max}}\leq\phi\leq\phi_{\text{max}}}{\text{min}}\sqrt{G(\phi-\eta_{n^*})} \left|H_M(\sin(\phi-\eta_{n^*}))\right|\geq \varepsilon.
\end{equation}


Furthermore, (\ref{we02}) can be equivalently rewritten as $N$ segmented functions, as follows

\begin{equation}\label{weX02}
\underset{\forall\phi\in \Omega_{n^*}, \forall {n^*}\in \mathcal{N}}{\text{min}} \enspace \sqrt{G(\phi-\eta_{n^*})} \left|H_M(\sin(\phi-\eta_{n^*}))\right|\geq \varepsilon. 
\end{equation}

Since the function $ \sqrt{G(\phi-\eta_{n^*})} \left|H_M(\sin(\phi-\eta_{n^*}))\right|$ is a periodic function, 
a sufficient condition for (\ref{weX02}) is
\begin{equation}\label{weX0x2}
 \underset{\forall\phi\in \Omega_{n^*}, n^*=0}{\text{min}}  \enspace  \sqrt{G(\phi-\eta_{n^*})} \left|H_M(\sin(\phi-\eta_{n^*}))\right|\geq \varepsilon.
\end{equation}

Thus, (\ref{weX0x2}) can be equivalently rewritten as
\begin{equation}\label{CC}
\begin{aligned}
\underset{|\phi|\leq0.5\arcsin\left(\frac{2}{M}\right)}{\text{min}} \sqrt{G(\phi)} \left|H_M(\sin\phi)\right|\geq \varepsilon.
\end{aligned}
\end{equation}


Since $\underset{x}{\min} \enspace f(x)g(x)\geq (\underset{x}{\min}\enspace f(x))(\underset{x}{\min}\enspace g(x))$ with $f(x)>0$ and $g(x)>0$, a sufficient condition for (\ref{CC}) is
\begin{equation}\label{CC2}
\begin{aligned}
 \underset{|\phi|\leq0.5\arcsin\bigl(\frac{2}{M}\bigl)}{\text{min}} G(\phi)  \geq \Bigl(\frac{\varepsilon}{\bigl|H_M\bigl(\sin\bigl(0.5\arcsin\left(\frac{2}{M}\right)\bigl)\bigl)\bigl|}\Bigl)^2.
\end{aligned}
\end{equation}


Therefore, under assumptions (i) and (ii) in Theorem \ref{lemmaX0001},  a sufficient condition to ensure (\ref{we01}) is
 \begin{equation}\label{CCX2}
\begin{aligned}
&G(0.5\phi_{\text{3dB}}) \geq \Bigl(\frac{\varepsilon}{\bigl|H_M\bigl(\sin\bigl(0.5\arcsin\left(\frac{2}{M}\right)\bigl)\bigl)\bigl|}\Bigl)^2,  \\
&\text{and } \phi_{\text{3dB}}\geq \arcsin\Bigl(\frac{2}{M}\Bigl ).
\end{aligned}
\end{equation}

Thus, the proof of Theorem \ref{lemmaX0001} is completed.

\section{ PROOF OF THEOREM \ref{AntennaPattern}}\label{2AP1}

 The antenna element pattern $G(\zeta)=10^{G_{\text{dB}}(\zeta)/10}$ in (\ref{G}) is equivalently expressed as
\begin{equation}\label{GX000}
\begin{aligned}
G(\zeta)=\Bigg\{
\begin{aligned}
&G(0)10^{-1.2\left(\frac{\zeta}{\phi_{\text{3dB}}}\right)^2}, |\zeta|\leq\sqrt{\frac{A_{\max}}{12}}\phi_{\text{3dB}}=\sqrt{2.5}\phi_{\text{3dB}},\\
&G(0)10^{-3} \quad\quad\quad\enspace,\sqrt{2.5}\phi_{\text{3dB}}<|\zeta|.
\end{aligned}
\end{aligned}
\end{equation}

For given the 3dB beamwidth $\phi_{\text{3dB}}$ and total power gain $G_{\text{sum}}$, we have
\begin{small}
\begin{equation}
\begin{aligned}
&G_{\text{sum}}=\int_{-\zeta_{\max}}^{\zeta_{\max}}G(\zeta)d\zeta\approx\int_{-\sqrt{2.5}\phi_{\text{3dB}}}^{\sqrt{2.5}\phi_{\text{3dB}}}G(0)10^{-1.2\left(\frac{\zeta}{\phi_{\text{3dB}}}\right)^2}d\zeta\\
&=2G(0)\phi_{\text{3dB}}\int_{0}^{\sqrt{2.5}}10^{-1.2x^2}dx=2G(0)\phi_{\text{3dB}}\int_{0}^{\sqrt{2.5}}e^{-a x^2}dx\\
&\approx G(0)\phi_{\text{3dB}}\sqrt{\pi/a}\cdot\text{erf}(\sqrt{2.5a})\approx1.066G(0)\phi_{\text{3dB}}.\\
\end{aligned}
\end{equation}
\end{small}%
where $a=1.2\ln 10$. Thus, the peak gain is given by
\begin{equation}
G(0)\approx\frac{G_{\text{sum}}}{1.066\phi_{\text{3dB}}}.
\end{equation}

Thus, the proof of Theorem \ref{AntennaPattern} is completed.

\section{ PROOF OF THEOREM \ref{lemmaX002}}\label{2DLE1}
\begin{figure}[htbp]
\centering
\includegraphics[width=0.66\linewidth]{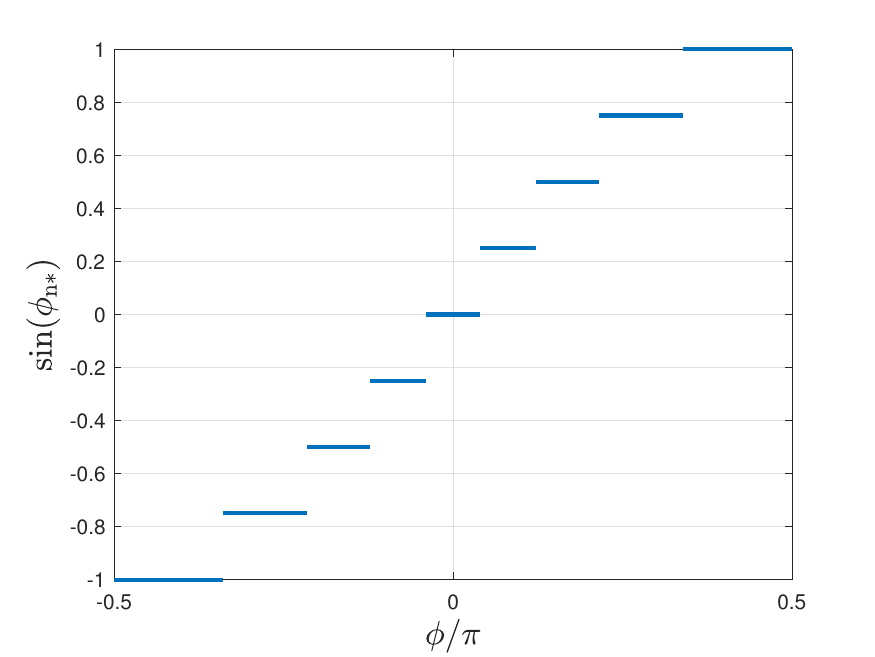}
	\caption{DFT codeword $\sin\phi_{n^*}$ with $n^*$ being the optimal solution to (\ref{DFTcoverage}) vs the signal direction  $\phi$.}
	\label{fenduan2}
\end{figure}

By substituting (\ref{steeringV1X}) into (\ref{DFTcoverage}), we have
\begin{equation}\label{DFTcoverage2}
f_{\text{ULA}}(\phi)
=M\sqrt{G_{\text{ULA}}(\phi)} \left|H_M(\sin\phi-\sin\phi_{n^*})\right|,
\end{equation}
where $\sin\phi_{n^*}=\frac{2n^*}{M}$ with $n^*=\bigl\lfloor\frac{\left\lfloor M\sin\phi\right\rfloor+1}{2}\bigl\rfloor$ being the optimal solution to (\ref{DFTcoverage}), as illustrated in Fig. \ref{fenduan2}. It is observed from Fig. \ref{fenduan2} that
when $\phi\in\Omega_{n^*}$ with $\Omega_{n^*}=\{\zeta:\text{max}\left(\arcsin\left(2(n^*-0.5)/M\right),-\phi_{\text{max}}\right)\leq\zeta<\text{min}\left(\arcsin\left(2(n^*+0.5)/M\right),\phi_{\text{max}}\right) \}$,
$\forall n^*\in\mathcal{N}'$, the index of selected DFT codeword $n^*$ remains unchanged. By substituting (\ref{DFTcoverage2}) and $ \varepsilon_0=M \varepsilon$ into (\ref{DFTcoverage1}), the antenna element pattern of the ULA $G_{\text{ULA}}(\zeta)$ needs to be designed to satisfy
\begin{equation}\label{DFTcoverage001}
\begin{aligned}
\underset{-\phi_{\text{max}}\leq\phi\leq\phi_{\text{max}}}{\text{min}}\sqrt{G_{\text{ULA}}(\phi)} \left|H_M(\sin\phi-\sin\phi_{n^*})\right|\geq \varepsilon.
\end{aligned}
\end{equation}

Thus, (\ref{DFTcoverage001}) can be equivalently rewritten as $N'$
segmented functions, as follows

\begin{equation}\label{DFTcoverage41}
\begin{aligned}
\underset{\forall \phi\in\Omega_{n^*},\forall n^*\in\mathcal{N'}}{\text{min}}\enspace \sqrt{G_{\text{ULA}}(\phi)} \left|H_M(\sin\phi-\sin\phi_{n^*})\right|\geq \varepsilon.
\end{aligned}
\end{equation}

Under the given assumption (iii) in Theorem \ref{lemmaX002}, 
a sufficient condition for (\ref{DFTcoverage41}) to hold can be derived when $\phi_{\text{3dB}}\geq2\phi_{\max}$, as follows
\begin{equation}\label{DFTcoverage401}
\begin{aligned}
\underset{\forall \phi\in\Omega_{n^*},n^*=\frac{N'-1}{2}}{\text{min}}\enspace \sqrt{G_{\text{ULA}}(\phi)} \left|H_M(\sin\phi-\sin\phi_{n^*})\right|\geq \varepsilon.
\end{aligned}
\end{equation}

Thus, (\ref{DFTcoverage401}) can be equivalently rewritten as
\begin{equation}\label{DFTcoverage04}
\begin{aligned}
 \underset{\arcsin\left(\frac{N'-2}{M}\right)\leq\phi\leq\phi_{\text{max}}}{\text{min}}\enspace \sqrt{G_{\text{ULA}}(\phi)} \Bigl|H_M\Bigl(\sin\phi-\frac{N'-1}{M}\Bigl)\Bigl|\geq \varepsilon.
\end{aligned}
\end{equation}

Since $\underset{x}{\min} \enspace f(x)g(x)\geq (\underset{x}{\min}\enspace f(x))(\underset{x}{\min}\enspace g(x))$ with $f(x)>0$ and $g(x)>0$,  a sufficient condition for (\ref{DFTcoverage04}) is

\begin{equation}\label{DFTcoverageXX04}
\begin{aligned}
 &\underset{\arcsin\left(\frac{N'-2}{M}\right)\leq\phi\leq\phi_{\text{max}}}{\text{min}}\enspace G_{\text{ULA}}(\phi) \geq\Bigl(\frac{\varepsilon}{\left|H_M(1/M)\right|}\Bigl)^2.
\end{aligned}
\end{equation}


Therefore, under assumption (iii) in Theorem \ref{lemmaX002}, a sufficient condition for (\ref{DFTcoverage1}) is
\begin{equation}\label{DFTcoverageP4}
G_{\text{ULA}}(0.5\phi_{\text{3dB}}^{\text{ULA}}) \geq \Bigl(\frac{\varepsilon}{\left|H_M(1/M)\right|}\Bigl)^2 \text{ with } \phi_{\text{3dB}}^{\text{ULA}}\geq2\phi_{\text{max}}.
\end{equation}

Thus, the proof of Theorem \ref{lemmaX002} is completed.

\end{appendices}

\bibliographystyle{IEEEtran}
\bibliography{./header_short,./bibliography1}

\end{document}